\documentclass[aps,prb,twocolumn,eqsecnum,amsmath,amssymb,floatfix]{revtex4}
\usepackage{tabularx}
\usepackage{bm}
\usepackage{euscript}
\usepackage{graphicx}
\usepackage{color}
\usepackage{amsfonts}
\usepackage{exscale}
\usepackage{amsbsy}

\begin{document}

\title{Fractional vortex lattice structures in spin triplet superconductors}

\author{Suk Bum Chung$^1$, Daniel F. Agterberg$^2$, and Eun-Ah Kim$^3$}
\affiliation{$^1$Department of Physics, Stanford University, Stanford, California 94305, USA\\
$^2$Department of Physics, University of Wisconsin-Milwaukee, Milwaukee, WI 53211\\
$^3$Department of Physics, Cornell University, Ithaca, NY 14853, USA}
\date{\today}

%%%% ALIASES and NEWCOMMAND%%%%%%%%%%%%%%%%%%%%%%%%%%%%%%%%%%%%%%%%%%

%%%%%%%%%%%%%%%%%%%%%%%%%%%%%%%%%%%%%%%%%%%%%%%%%%%%%%%%%%%%%%%%%%%%%%%

\begin{abstract}

Motivated by recent interest in spin-triplet superconductors, we
investigate the vortex lattice structures for this class of 
unconventional superconductors. We discuss how the order parameter 
symmetry can give rise to U(1)$\times$U(1) symmetry in the same 
sense as in spinor condensates, making half-quantum vortices (HQVs) 
topologically stable.We then calculate the vortex lattice structure 
of HQVs, with particular attention on the roles of the crystalline
lattice, the Zeeman coupling and Meissner screening, all absent in 
spinor condensates. Finally, we consider how spin.orbit coupling 
leads to a breakdown of the U(1)$\times$U(1) symmetry in free 
energy and whether the HQV lattice survives this symmetry breaking. 
As examples, we examine simpler spin-triplet models proposed in the 
context of Na$_x$CoO$_2 \cdot y$H$_2$O and Bechgaard salts, as well 
as the better known and more complex model for Sr$_2$RuO$_4$.

\end{abstract}

\maketitle

\section{Introduction}
A half quantum vortices (HQV) with vorticity $h/4e$, which is half
that of usual Abrikosov vortex with vorticity $\Phi_0\equiv h/2e$,
presents an exciting example of fractionalized topological defects. 
Quantization of collective topological defects provides clear cut 
access to the nature of the ground state. For instance, the 
vorticity $h/2e$ of Abrikosov vortex in a type II superconductor 
clearly shows that the circulation associated with the vortex is 
that of charge $2e$ Cooper pairs. Since the vorticity in a 
condensate is determined by the requirement of single-valuedness of 
the order parameter describing the condensate, ``fractionalization 
of vorticity'' is possible with a multi-component order parameter 
when different components are allowed to wind separately. For 
instance, in a triplet superconductor, the additional Cooper pair 
spin degree of freedom can be free to rotate in plane 
\cite{PhysRevLett.55.1184,PhysRevLett.89.067001,chung:197002} giving 
rise to additional U(1) symmetry and an associated spin winding 
number; cases where vortex fractionalization is due to the 
U(1)$\times$U(1) symmetry of different physical origin has also been 
studied \cite{PhysRevLett.94.137001}. Hence observation of 
fractionalization of vortices can serve as an indicator of the 
structure of the order parameter in a given condensate. In addition, 
the recent proposals predicting non-Abelian fractional statistics for 
the composite of a HQV and the Majorana fermions bound at its core in 
the chiral triplet superconductors brought in recent rise in the 
attention and interest to the possibility of HQV's in triplet 
superconductors \cite{PhysRevB.61.10267,PhysRevLett.86.268,PhysRevB.70.205338,PhysRevB.73.014505}. 
This type of non-Abelian statistics was first studied for quasiholes 
in the spin-polarized $\nu = 5/2$ quantum Hall state \cite{moore-read,nayak-wilczek}.
Therefore, if we want to obtain the same statistics for vortices in a 
spinful superconductor, the vortices should be HQVs so that there 
would be phase winding only for a single component.

However, while there are a number of candidate triplet superconductors 
such as the single layer ruthenate Sr$_2$RuO$4$ 
\cite{Maeno:1994fk,RevModPhys.75.657}, the cobaltate 
Na$_x$CoO$_2 \cdot y$H$_2$O \cite{Nature.Physics.1.91} and organic 
superconductors \cite{organic}, HQV's have never been observed in bulk
systems, in line with the energetic stability issues raised by two of 
us in Ref.~\onlinecite{chung:197002}. It was pointed out in
Ref.~\onlinecite{chung:197002} due to the absence of screening for the
spin supercurrent circulation required for HQV in triplet 
superconductors, HQV's can be energetically unstable in bulk samples 
towards combining into full Abrikosov vortices despite their advantage 
in magnetic energy. Related considerations have appeared in the context 
of spin-triplet superconductivity in UPt$_3$ by Zhitomirsky 
\cite{Zhitomirskii_UPt3}.

The main motivation of this work is to investigate the possibility
of using high enough fields to generate a HQV lattice in triplet
superconductors where the vortex lattice serves two purposes at
once: 1) stabilizing HQV's at finite separation 2) providing an
unambiguous signature of its formation (halving of the vortex
lattice unit cell). Experiments have already determined the vortex 
lattice structure successfully at low fields in Sr$_2$RuO$_4$
\cite{Nature.396.242, PhysRevLett.84.6094} and the observed
square lattice geometry was consistent with the theoretical
prediction by one of the present authors based on a chiral triplet
order parameter in Ref.~\onlinecite{PhysRevB.58.14484}. However,
Ref.~\onlinecite{PhysRevB.58.14484} considered the limit of strong
spin-orbit coupling, which leads to vortex lattices of full
quantum Abrikosov vortices. Recently,  measurements of the Knight
shift for the field along the $c$-axis \cite{PRL.93.167004} as well
as ARPES data \cite{haverkort:026406,liu:026408} indicate that that 
spin-orbit coupling is perhaps not so strong. Therefore, in this 
paper, we extend the studies of Ref.~\onlinecite{PhysRevB.58.14484} 
to allow for weak spin-orbit coupling, leading to the possibility 
of HQV lattices for fields along the $c$-axis. Additionally, the 
organic superconductor (TMTSF)$_2$ClO$_4$ naturally has weak 
spin-orbit coupling and Knight shift measurements provide evidence 
for a spin-triplet state at high magnetic fields \cite{organic}, 
which is precisely the situation we consider here. We also provide 
an analysis of this case and the closely related case for cobaltate
spin-triplet superconductors \cite{Nature.Physics.1.91}.

In this paper, we study the energetics of different vortex 
lattice configurations. The key additional physical ingredient is 
the U(1) spin-rotational invariance of the Cooper pairs that
arises in a magnetic field. This generically leads to two 
different species of fractional vortices whose fractional fluxes 
sum to $\Phi_0$. When stable, these fractional vortices form 
interlacing lattices analogous to vortex.antivortex lattice 
configurations proposed by Refs.~\textcite{PhysRevLett.71.2138, PhysRevLett.71.2142} 
in the context of a two-dimensional (2D) superfluid and the 
configuration in two-component Bose condensates proposed by 
Ref.~\textcite{PhysRevLett.88.180403}.

The rest of the paper consists of the following. In section
\ref{sec:OP}, we give a pedagogical introduction to the symmetry
properties of triplet OP. In particular, we will show how
U(1)$\times$U(1) symmetry can arise in the OP of such systems. In
section \ref{sec:GibbsFree} we discuss the form of Gibbs free
energy that is allowed by various symmetries in the problem when
spin-orbit coupling is not included. In section \ref{sec:G} we
provide the general theoretical framework for the vortex lattice
phases. In section \ref{sec:LLL} we show that the lowest Landau
level solution often provides an adequate description and we
discuss this solution for the lattice of HQV's.  In section
\ref{sec:SO} we consider the effect of U(1)$\times$U(1) symmetry
breaking driven by spin-orbit coupling. In section
\ref{sec:observation} we lay out predictions for how to detect the
proposed HQV lattice structures and we conclude with a summary and
outlook in section \ref{sec:conclusion}.

\section{The triplet order parameter}
\label{sec:OP}

The order parameter of a triplet superconductor takes a matrix 
form in the spin space\cite{RevModPhys.75.657,RevModPhys.63.239}:
\begin{equation}
\hat{\Delta} ({\bf k}) \!=\! \left [
\begin{array}{cc}
  \Delta_{\uparrow\uparrow}({\bf k}) & \Delta_{\uparrow\downarrow}({\bf k}) \\
  \Delta_{\downarrow\uparrow}({\bf k}) & \Delta_{\downarrow\downarrow}({\bf k})\\
\end{array}
\right ] \!\equiv\! \left[
  \begin{array}{cc}
  -d_x + id_y & d_z \\
  d_z & d_x + id_y \\
\end{array}
\right ], \label{EQ:dVec}
\end{equation}
where the spin quantization axis is along the $z$ direction. The
triplet pairing requires
$\Delta_{\uparrow\downarrow}\!=\!\Delta_{\downarrow\uparrow}$ and
a set of three complex functions of ${\bf k}$, namely 
$(d_x({\bf k}),d_y({\bf k}), d_z({\bf k}))$, were introduced to 
parameterize the gap matrix. When the three functions are 
collectively represented using a vector notation, the ``unit vector'' 
${\bf\hat d}({\bf k})$ represents the symmetry direction (zero 
projection direction) with respect to the rotation of Cooper pair 
spin. In the presence of the sufficiently high field along the 
$c$-axis, Zeeman splitting between electrons with opposite spins 
prohibits pairing, leading to 
$\Delta_{\uparrow\downarrow}= \Delta_{\downarrow\uparrow}=0$. 
In the $\bf{d}$-vector notation, this implies that the 
$\bf{d}$-vector lies in-plane (perpendicular to the applied field). 
In the rest of the paper, we assume that the field is sufficiently 
large so that this is the case. In the context of strontium 
ruthenate, our results apply for the field along the $c$-axis (this 
is also true for the cobaltates when we are discussing spin-orbit 
coupling in hexagonal systems). For organic and cobaltate 
superconductors, our results apply for the field along any two-fold 
or higher symmetry axis of symmetry.

For non-chiral triplet order parameter symmetry, which is expected
of the  cobaltate Na$_x$CoO$_2 \cdot y$H$_2$O 
\cite{Nature.Physics.1.91} and organic superconductors \cite{organic}, 
the spin pairing gap matrix takes the form
\begin{equation}
\hat{\Delta}({\bf k})=f({\bf k})\left[
 \begin{array}{cc}
  \Delta_{\uparrow\uparrow}& 0 \\
  0 & \Delta_{\downarrow\downarrow}
  \end{array}
\right] \label{EQ:OP-nonchiral}
\end{equation}
where the function $f({\bf k})$ depends on the specific odd
angular momentum channel. The key simplifying feature is the the
orbital dependence is described by a one-dimensional 
representation encoded by $f({\bf k})$. There has been suggestions
that the cobaltate Na$_x$CoO$_2 \cdot y$H$_2$O has a spin-triplet
pairing through $f$-wave channel \cite{Nature.Physics.1.91},
although data from the Knight shift experiments remain
controversial \cite{PhysRevB.73.180503, JPSJ.77.073702}. In this
case a common choice is $f({\bf k}) = k_x(k_x^2-3k_y^2)$. However,
the precise form of $f({\bf k})$ is not needed for our results.
For the organic superconductor (TMTSF)$_2$ClO$_4$, there is a
strong case that the system becomes a triplet superconductor under
sufficient field $H\gtrsim $20kOe. In this case, there are many
proposals for $f({\bf k})$. However, again, the specific form is
not needed for our results.

For chiral order parameter symmetry expected of the Sr$_2$RuO$_4$,
with $\bf{d}$ in the basal plane and no spin-orbit coupling, the
order parameter has four complex degrees of freedom:
\begin{equation}
\hat{\Delta}({\bf k})=
\sum_{\sigma=\pm}\tilde{f}(k_\sigma,k_z)\left[
 \begin{array}{cc}
  \Delta_{\uparrow\uparrow,\sigma}& 0 \\
  0 & \Delta_{\downarrow\downarrow,\sigma}
  \end{array}
\right], \label{EQ:OP-chiral}
\end{equation}
where the function $\tilde{f}$, like the non-chiral case discussed
above, depends on the specific odd angular momentum channel and
$k_\sigma = k_x + i\sigma k_y$. This is equivalent to
\begin{equation}
{\bf d} ({\bf k}) = \Delta_+ {\bf {\hat d}}_+ \exp (i
n\varphi_{\bf{\hat k}})+\Delta_- {\bf {\hat d}}_- \exp (-i
n\varphi_{\bf{\hat k}}), \label{EQ:OPp+ip}
\end{equation}
where $n$ is an integer ($n=1$ for $p$-wave, $n=3$ for $f$-wave)
and $\varphi_{\bf{\hat k}}$ is the azimuthal angle associated with
a unit vector $\hat{\bf k}$ in the 2D plane (assuming quasi-2D
setting with the angular momentum along the $c$-axis: ${\bf \hat
l} = {\bf \hat z}$). Although $\Delta_- = 0$ for homogeneous
chiral superconductor, we will show that $\Delta_- \neq 0$ often
plays an important role in describing the vortex lattice structure
of chiral superconductor.

Eqs.(\ref{EQ:OP-nonchiral}) and (\ref{EQ:OPp+ip})  clearly shows
that under these circumstances, the order parameter symmetry takes
the $U(1)\times U(1)$ form, which can allow for HQV's with $h/4e$
vorticity associated with $\pi$ orbital phase winding and $\pi$
${\bf d}$-vector winding.

\section{The Gibbs free energy}
\label{sec:GibbsFree}

In order to identify stable vortex type and the lattice structure
itself, we start with the Gibbs free energy including all the
terms allowed by symmetry up to quartic order. As usual, the
quartic  terms determine the vortex lattice structure. Due to
additional spin degrees of freedom, the full expression for the
Gibbs free energy involves a number of additional terms compared
to singlet superconductor case and it is instructive to consider
different contributions separately:
\begin{equation}
f = f_{mag} + f^{(2)}_0 + f^{(2)}_{Z} + f^{(2)}_{in} +
f^{(2)}_{SO} +  f^{(4)}_{hom} + f^{(4)}_{in}. \label{EQ:freeTot}
\end{equation}
where   $f_{mag} = h^2/8\pi - hH/4\pi$  is the magnetic energy
(the  field $h$ is the sum of the external field $H$ and the
screening field), the superscript (2) indicates terms quadratic in
OP and (4) quartic in OP

Other than the conventional quadratic term  $f^{(2)}_0$:
\begin{equation}
f^{(2)}_0 = -\alpha \sum_i |\Delta_i|^2 \label{EQ:f0}
\end{equation}
the remaining quadratic terms in Eq.\eqref{EQ:freeTot} are
consequences of additional spin degree of freedom for the triplet
superconductors. The Zeeman coupling term
\begin{equation}
f^{(2)}_{Z} = -\tilde{\kappa} h (|\Delta_{\uparrow\!\uparrow}|^2 -
|\Delta_{\downarrow\!\downarrow}|^2),
\end{equation}
plays an important role for the HQV vortex lattice by introducing
a slight spin-polarization. This slight spin-polarization gives
rise to two phase transitions as in the case of the A$_1$/A$_2$ phase of
$^3$He\cite{RevModPhys.47.331, VollhardtHe3}. The inhomogeneous
part of the quadratic terms $f^{(2)}_{in}$ are of the form
\begin{equation}
K_{ij;kl} (D_i \Delta_k) (D_j \Delta_l)^* + {\rm c.c.},
\end{equation}
where $D_i = \nabla_i + (2\pi i/\Phi_0)A_i$. For these terms, we
require rotational invariance up to the lattice symmetry with
respect to orbital degrees of freedom only, which means that we
require invariance with respect to rotating $D_i$'s and the
orbital component of $\Delta_i$'s. The complex structure of
$f^{(2)}_{in}$ can result in the condensate wave function of a
different form than that of conventional SC. $f^{(2)}_{SO}$ is the
quadratic spin-orbit coupling term assuming the spin-orbit
coupling to be small and is discussed in Sec.~\ref{sec:SO}. In the
presence of spin-orbit coupling the free energy have to be
invariant under the combined discrete rotation of the orbital and
spin degree of freedom specific for the given lattice symmetry.
For lattices with orthogonal or tetragonal symmetry, spin-orbit
coupling may reduce the symmetry of the Gibbs free energy and tend 
to suppress HQV formation by introducing a length scale beyond 
which the HQV's cannot exist (this length scale diverges as the 
spin-orbit coupling vanishes). This implies that the vortex lattice 
spacing must be less than this length scale for the HQV lattice to 
appear. However, we show that for spin-triplet hexagonal materials 
(specifically the two-dimensional $\Gamma^-_6$ and $\Gamma^-_5$ 
representations in the notation of Sigrist and Ueda 
\cite{RevModPhys.63.239}), even large spin-orbit coupling still 
allows for the existence of a fractional vortex lattice. This 
consideration may apply to Na$_x$CoO$_2\cdot$yH$_2$O.

Among the quartic terms, $f^{(4)}_{hom}$ represents the usual set
of homogeneous terms. As is shown in Appendix 
\ref{app:weakcoupling}, certain quartic terms vanish in the 
weak-coupling theory. The terms that vanish are those that lift 
the energy degeneracy between the full quantum vortex (QV) and the 
HQV lattice. For this reason, we also include the inhomogeneous
quartic term $f^{(4)}_{in}$. This term accounts for the difference
between the spin phase stiffness $\rho_{sp}$ and the overall phase
stiffness $\rho_s$. Not only does this difference play an
important role in the stability of isolated HQV's as it was shown
in Ref.~\onlinecite{chung:197002} it plays the role of tuning
parameter for the vortex lattice structure.

With multiple systems in mind, we consider contributions to the 
Gibbs free energy specific for non-chiral and chiral 
superconductors respectively.

\subsection{Non-chiral Triplet Superconductor}

As mentioned earlier, here we assume the orbital dependence of 
the gap function to be the same for all spin-triplet components. 
Formally, this means that the orbital degree of freedom belongs to 
a one-dimensional irreducible representation of the point group. 
We apply our analysis to materials that have orthorhombic, 
tetragonal, or hexagonal point groups. One relevant example is a
non-chiral triplet $f$-wave superconductor with hexagonal
symmetry, which has been proposed in the context of the the
cobaltates Na$_x$CoO$_2\cdot y$H$_2$O; in this case  
$f({\bf k}) = k_x(k_x^2-3k_y^2)$ from Eq.\eqref{EQ:OP-nonchiral}. 
When the ${\bf d}$-vector lies in the xy plane, the inclusion of 
spin-orbit coupling implies that formally this order parameter 
belongs to the $\Gamma_6^-$ representation of the hexagonal point 
group (the consequences of spin-orbit coupling for this 
representation is discussed in more detail in Section 
\ref{sec:SO}).

With the in-plane spin rotational invariance, the relevant free 
energy within the assumptions stated above is given by
\begin{align}
&f^{(2)}_{in} = \sum_{i=x,y,z} K_i(|D_i\Delta_{\uparrow\uparrow}|^2+
|D_i\Delta_{\downarrow\!\downarrow}|^2),\label{EQ:f2in-NX}\\
&f^{(4)}_{hom}=\beta_1(\sum_{i}|\Delta_{i}|^2)^2+\beta_2|\Delta_{\uparrow\!\uparrow}|^2
|\Delta_{\downarrow\!\downarrow}|^2,\label{EQ:f2hom-NX}\\
&f^{(4)}_{in} =
\gamma[\Delta_{\uparrow\!\uparrow}^*\Delta_{\downarrow\!\downarrow}(\textbf{D}_{\perp}\Delta_{\uparrow\!\uparrow})\!\cdot\!(\textbf{D}_{\perp}\Delta_{\downarrow\!\downarrow})^*+{\rm c.c}],\label{EQ:f4-NX}
\end{align}
For tetragonal and hexagonal point groups $K_x=K_y$ while for
orthorhombic point groups, $K_x\ne K_y$. For the high field limit
we are considering, it is possible to re-scale lengths in two
directions perpendicular to applied field such that
$\tilde{K}_i=\tilde{K}_j$  for $i\ne j$ (where $\tilde{K}_i$
refers to the new coefficient in the re-scaled coordinates). We
will therefore ignore the difference between the $K_i$ and assume
that for orthorhombic point groups we are working in re-scaled
co-ordinates. The term $f_{in}^{(4)}$ is not the most general such
term allowed by symmetry. However, it is this term that allows the
GL theory to give the same physics as in 
Ref.~\onlinecite{chung:197002}. Indeed, one can gain more insight
into the vortex lattice solutions that minimizes Eq.\eqref{EQ:f0}
and Eqs.(\ref{EQ:f2hom-NX}-\ref{EQ:f4-NX}) by relating the
coefficient of the  inhomogeneous quartic term $\gamma$ to the
stiffness ratio $\rho_{sp}<\rho_s$ which controls the energetic
stability of a pair of HQV's\cite{chung:197002}. Within the London
approximation,  the gradient terms in Eq.\eqref{EQ:f2in-NX} and
Eq.\eqref{EQ:f4-NX} amounts to phase bending energy which will be
proportional to $(\rho_{s}+\rho_{sp})$ and $(\rho_{s}-\rho_{sp})$
respectively for $\Delta_{\uparrow\!\uparrow}$ and
$|\Delta_{\downarrow\!\downarrow}|$. Combining 
Eq.\eqref{EQ:f2in-NX} and Eq.\eqref{EQ:f4-NX} with the homogeneous
solution 
$|\Delta_{\uparrow\!\uparrow}|^2=|\Delta_{\downarrow\!\downarrow}|^2 = \alpha/(\beta_1-\beta_2)$, 
we obtain the following relation between $\gamma$ and 
$\rho_{sp}/\rho_s$:
\begin{equation}
\gamma = \frac{K_1(\beta_1-\beta_2)}{\alpha}
\frac{1-\rho_{sp}/\rho_s}{1+\rho_{sp}/\rho_s}.
\end{equation}
Hence $\gamma >0$ would imply stability of HQV's and double
transitions into two possible vortex phases: a lattice of ordinary 
Abrikosov vortices and a lattice of HQV's. This transition is 
determined by the $\beta_2$ term of Eq.~(\ref{EQ:f2hom-NX}) and
the $\gamma$ term of Eq.~(\ref{EQ:f4-NX}).

\subsection{Chiral Triplet Superconductor}
With  the ruthenate Sr$_2$RuO$_4$ in mind, we consider a chiral 
triplet $p$-wave superconductor with square symmetry for which 
$\tilde{f}(k_\sigma) = k_x + i\sigma k_y$ in the 
Eq.\eqref{EQ:OP-chiral} with
\begin{equation}
\hat{\Delta}({\bf k})=  \sum_{\sigma=\pm}(k_x + i\sigma k_y)\left[
 \begin{array}{cc}
  \Delta_{\uparrow\uparrow,\sigma}& 0 \\
  0 & \Delta_{\downarrow\downarrow,\sigma}
  \end{array}
\right], \label{EQ:spinChiral1}
\end{equation}
where $\Delta_{s,\sigma}$
($s =\uparrow\!\uparrow,\downarrow\!\downarrow$ and $\sigma=\pm$)
form expansion parameters for the Gibbs free energy. In terms of the
$\bf{d}$-vector notation 
${\bf d} \equiv {\bf \hat{x}}(\eta_{xx} k_x + \eta_{xy} k_y) + {\bf \hat{y}}(\eta_{yx} k_x + \eta_{yy} k_y)$,
\begin{eqnarray}
\Delta_{\uparrow\!\uparrow,+} = &-& (\eta_{xx} - i\eta_{xy} - i\eta_{yx} - \eta_{yy})/2,\nonumber\\
\Delta_{\uparrow\!\uparrow,-} = &-& (\eta_{xx} + i\eta_{xy} - i\eta_{yx} + \eta_{yy})/2,\nonumber\\
\Delta_{\downarrow\!\downarrow,+} = &\,& (\eta_{xx} - i\eta_{xy} + i\eta_{yx} + \eta_{yy})/2,\nonumber\\
\Delta_{\downarrow\!\downarrow,-} = &\,& (\eta_{xx} + i\eta_{xy} +
i\eta_{yx} - \eta_{yy})/2. \label{EQ:spinChiral2}
\end{eqnarray}

Formally, without spin-orbit coupling,  this order parameter is a
direct product of a $E_u$ orbital representation of the tetragonal
point group and the in-plane vector representation for spin
rotations. When spin-orbit is included the order parameter 
contains the four different one-dimensional representations of the
tetragonal point group. In the case without spin-orbit coupling,
the relevant free energy for this representation can be
constructed using the known free energy for the $E_u$ 
representation \cite{RevModPhys.63.239}, we list below
$f^{(2)}_{in}$, $f^{(4)}_{hom}$ and $f^{(4)}_{in}$. Before listing
$f^{(2)}_{in}$, we note that this free energy term should respect
the $C_4$ symmetry on the $xy$ plane only for the orbital degrees
of freedom:
\begin{equation}
(D_x, D_y, \Delta_{s,+}, \Delta_{s,-}) \to (D_y, -D_x,
i\Delta_{s,+}, -i\Delta_{s,-}).
\end{equation}
However, for simplicity, we consider cylindrical symmetry in the 
orbital degrees of freedom, this does not significantly alter the 
arguments below. This symmetry gives us 
$f^{(2)}_{in} = \sum_s f^{(2,s)}_{in}$ where
\begin{eqnarray}
f^{(2,s)}_{in} &=& K_1(|{\bf D}\Delta_{s,+}|^2 + |{\bf D}\Delta_{s,-}|^2)\nonumber\\
&+& K_2[\!\{\!(D_x \Delta_{s,+})\!(D_x \Delta_{s,-})^*\!-\!(D_y \Delta_{s,+})\!(D_y \Delta_{s,-})^*\!\}\!/2\nonumber\\
&+& \!\{\!(D_x \Delta_{s,-})\!(D_x \Delta_{s,+})^*\!-\!(D_y \Delta_{s,-})\!(D_y \Delta_{s,+})^*\!\}\!/2\nonumber\\
&+& i\!\{\!(D_x \Delta_{s,-})\!(D_y \Delta_{s,+})^*\!+\!(D_y \Delta_{s,-})\!(D_x \Delta_{s,+})^*\!\}\!/2\nonumber\\
&-& i\!\{\!(D_x \Delta_{s,+})\!(D_y \Delta_{s,-})^*\!+\!(D_y \Delta_{s,+})\!(D_x \Delta_{s,-})^*\!\}\!/2]\nonumber\\
&+& K_4 (|D_z\Delta_{s,+}|^2 + |D_z\Delta_{s,-}|^2).
\label{EQ:quadDeriv2}
\end{eqnarray}
In addition, the following term is also allowed by symmetry
\begin{equation}
\delta K \frac{2\pi}{\Phi_0}h \sum_s (-|\Delta_{s+}|^2 + |\Delta_{s-}|^2),
\label{EQ:chiralSplit}
\end{equation}
and it stabilizes this in-plane chiral phase for strong enough 
magnetic field (note the similarity to the Zeeman term for the 
condensate spin degrees of freedom). As for the homogeneous
quartic terms,
\begin{eqnarray}
f^{(4)}_{hom} &=& \sum_s [\beta_1(|\Delta_{s+}|^4+|\Delta_{s-}|^4)/2 + \beta'_1|\Delta_{s+}|^2 |\Delta_{s-}|^2]\nonumber\\
&-&\sum_{\sigma=\pm}
(\beta_2|\Delta_{\uparrow\!\uparrow,\sigma}|^2
|\Delta_{\downarrow\!\downarrow,\sigma}|^2 + \beta'_2
|\Delta_{\uparrow\!\uparrow,\sigma}|^2
|\Delta_{\downarrow\downarrow,-\sigma}|^2)\nonumber\\
&-&\beta_3[(\Delta_{\uparrow\!\uparrow,+}\Delta_{\downarrow\!\downarrow,-})
(\Delta_{\uparrow\!\uparrow,-}\Delta_{\downarrow\!\downarrow,+})^*
+ {\rm c.c.}]. \label{EQ:int1}
\end{eqnarray}
$\beta_2$, $\beta'_2$ and $\beta_3$ terms originate from
interaction between spin up-up pairs and down-down pairs. Again,
for simplicity, we written the free energy in the limit of a
cylindrical Fermi surface. Lastly, we have
\begin{eqnarray}
f^{(4)}_{in} &=& \gamma \sum_{\sigma=\pm} [\Delta^*_{\uparrow\uparrow,\sigma} \Delta_{\downarrow\downarrow,\sigma} ({\bm D}\Delta_{\uparrow\uparrow,\sigma})\cdot ({\bm D} \Delta^*_{\downarrow\downarrow,\sigma}) + {\rm c.c.}]\nonumber\\
&+& \gamma' \sum_{\sigma=\pm} [\Delta^*_{\uparrow\uparrow,\sigma} \Delta_{\downarrow\downarrow,-\sigma} ({\bm D}\Delta_{\uparrow\uparrow,\sigma})\cdot ({\bm D} \Delta^*_{\downarrow\downarrow,-\sigma}) + {\rm c.c.}].\nonumber\\
\label{EQ:int2}
\end{eqnarray}
Note that the form of Eq.~(\ref{EQ:int2}) is consistent with the
form of the interaction terms in Eq.~(\ref{EQ:int1}). Again, this
is not the most general term allowed by symmetry, but it is the
minimal term that captures the physics in the London limit
described in Ref.~\onlinecite{chung:197002}.

\section{Determining the vortex lattice structure - general considerations}
\label{sec:G}

We consider the vortex lattice phases near the upper critical 
field to map out the stability condition for HQV lattice 
phases. As usual, the first step towards determining the vortex 
lattice structure is to identify the eigenstates of the 
linearized Ginzburg-Landau (GL) equations. In order  to obtain a 
linearized GL equation we take a variation of the quadratic 
terms in the free energy, for example:
\begin{eqnarray}
f^{(2)}_0 + f^{(2)}_{in} + f^{(2)}_Z &=& -\alpha \sum_j
|\Delta_j|^2 - \tilde{\kappa} h (|\Delta_{\uparrow\!\uparrow}|^2 -
|\Delta_{\downarrow\!\downarrow}|^2)\nonumber\\&+& [K_{jk;lm} (D_j \Delta_l) (D_k \Delta_m)^* + {\rm c.c.}]\nonumber\\
\end{eqnarray}
with respect to one component of the order parameter 
$\Delta_{s}^*$. This gives an equation of the form
\begin{equation}
\alpha \Delta_s = K_{kl;ss'}^* D_k D_l \Delta_s' -
\tilde{\kappa}H\Delta_s \label{EQ:genLinGL}
\end{equation}
(note that we are ignoring the difference between $h$ and $H$ in
this approximation). Since the gradient terms in Eq.\eqref{EQ:genLinGL}
cannot in general be reduced into a $D_x^2 + D_y^2$ form, the lowest
Landau level wave functions are not sufficient for  calculating the
condensate wave function in general. However, the solution of this
equation can still be expressed in terms of Landau level wave functions:
\begin{eqnarray}
\phi_n ({\bf r}) &=& [2^n \pi^{1/2} (n!)]^{-1/2}\nonumber\\ &\times& \sum_m q_m e^{i k_m x'} e^{-(y'-k_m)^2/2} H_n (y'-k_m),\nonumber\\
\end{eqnarray}
where $H_n$ is the Hermite polynomial of n$^{th}$ order, $x'$ and
$y'$ are $x$, $y$ coordinates in the unit of the magnetic length
$l = (\Phi_0/2\pi H)^{1/2}$. This is because $D_i$'s can be expressed as
a linear combination of the raising and lowering operators of the
Landau levels $\Pi_\pm$, since $\Pi_\pm = l(D_x \pm
iD_y)/\sqrt{2}$.

This wave function describes a vortex lattice when $|\phi_n ({\bf r})|$
is periodic in the lattice vectors
${\bf a_1} = al (1,0)$ and ${\bf a_2} = bl (\cos \theta,\sin \theta)$,
and $\phi_n ({\bf r})$ vanish at $m_1 {\bf a_1} + m_2 {\bf a_1}$
when $m_1$ and $m_2$ are integers. This requires
\begin{eqnarray}
k_m &=& 2\pi (m-1/2)/a = (m-1/2)\sqrt{2\pi \sigma}\nonumber\\
q_m &=&  e^{i\pi m (\varsigma + 1 - m\varsigma)},
\end{eqnarray}
where $\sigma = (b/a)\sin \theta$ and $\varsigma = (b/a)\cos\theta$. 
Note that we used the flux quantization condition 
$ab \sin\theta = 2\pi$. For a lattice of HQV's we need to consider 
a second lattice that is translated by 
$l{\bm \tau} = l (\tau_x, \tau_y)$ with respect to the first lattice.
For the wave function of this lattice, we can use \cite{superRev}
\begin{equation}
\tilde{\phi}_n ({\bf r}) = e^{i \tau_y x} \phi_n ({\bf r} - {\bm
\tau}),
\end{equation}
the phase factor being chosen so that 
$\Pi_- \tilde{\phi}_0 ({\bf r}) = 0$. This latter condition ensures
that the translated eigenstates have the same gauge as the 
untranslated eigenstates.

The formalism considered here gives us not only the energy due to 
interaction between vortices but also the core energy of vortices 
as well. This is because our vortex lattice wave function gives 
full description of the core regions. From the linearized 
GL equation we use here, a full quantum vortex is merely two HQV's 
of opposite spins coinciding at a same point. This means that the 
full quantum vortex core energy, if we ignore the cross term between 
two spin components in $f^{(4)}_{hom}$, is approximately twice the 
core energy of a HQV. If the HQV core energy is actually larger than 
this, that would make stabilization of the HQV more difficult, i.e., 
the largest allowed value for $\rho_{sp}/\rho_s$ for the HQV lattice 
would be smaller than what we obtain through the formalism used here.

The lattice structure can be determined by finding
$(\sigma,\varsigma, {\bm \tau})$ that minimize the free energy
expectation value. For this, we first set the amplitude of OP to
minimize the energy for given $(\sigma,\varsigma, {\bm \tau})$
(the amplitude depends on $H_{c2}-H$), and then
compare energy for different values of  $(\sigma,\varsigma, {\bm
\tau})$. To determine these structures we will need to take the
spatial integral of the product of four Landau level
wavefunctions. We have computed these integrals in the
Appendix~\ref{app:corr}.

\section{Lowest Landau level solution}
\label{sec:LLL}

In the bulk of this section we provide detailed analysis of the lowest
Landau level solution for the non-chiral triplet superconductors and
briefly comment on the chiral case in subsection \ref{subsec:chiral}.
In this case the relevant free energy (for orthorhombic, tetragonal,
and hexagonal materials) is
\begin{eqnarray}
f &=& \sum_{s=\uparrow\!\uparrow, \downarrow\!\downarrow}
\left[-\alpha|\Delta_s|^2 + \beta_1 |\Delta_s|^4/2 + \left(\sum_{i=x,y,z} K_i|D_i\Delta_s|^2\right)\right]\nonumber\\
&-& \beta_2\!|\Delta_{\uparrow\!\uparrow}|^2 |\Delta_{\downarrow\!\downarrow}|^2
- \tilde{\kappa} h (|\Delta_{\uparrow\!\uparrow}|^2 - |\Delta_{\downarrow\!\downarrow}|^2)\nonumber\\
&+& \gamma[\Delta_{\uparrow\!\uparrow}^*\Delta_{\downarrow\!\downarrow}
(\textbf{D}\Delta_{\uparrow\!\uparrow})\!\cdot\!(\textbf{D}\Delta_{\uparrow\!\uparrow})^*+{\rm c.c}]
+ \frac{h^2}{8\pi} - \frac{H h}{4\pi}.\nonumber\\
\label{EQ:freeLLL1}
\end{eqnarray}
As mentioned before, we assume that we have re-scaled lengths so
that we can take $K_z=K_x=K_y=K$. First look at the upper critical
field problem. The linearized GL equation is:
\begin{equation}
\frac{\alpha l^2}{K} \Delta_{s}= \left(1+2\Pi_+\Pi_- -s\frac{Hl^2}{K}\tilde{\kappa}\right) \Delta_{s}.
\end{equation}
The largest $H_{c2}$ occurs when the $\Delta_s$ are in the lowest
Landau level. This leads to two possible values for the upper critical
field,
\begin{equation}
H_{c2}^{\pm}=\frac{\alpha \Phi_0}{2\pi(K\pm Hl^2\tilde{\kappa})}
\end{equation}
We assume that $\tilde{\kappa}>0$ so that $\Delta_{\uparrow\!\uparrow}$ 
has the larger $H_{c2}$. We do not assume that the splitting between 
these two critical fields is large since it is given by the small 
Zeeman term $\tilde{\kappa}$.

Now consider the expectation value of $f$ in terms of the Landau level
wave functions. Applying Eq.~(\ref{EQ:genLinGL}) to the first two 
terms of Eq.~(\ref{EQ:freeLLL1}) gives the lowest Landau level solutions 
$\Delta_{\uparrow\!\uparrow} = C_{\uparrow\!\uparrow} \phi_0$ and
$\Delta_{\downarrow\!\downarrow} = C_{\downarrow\!\downarrow} e^{i 2\alpha} \tilde{\phi}_0$.
Inserting this solution as determined at $H=H_{c2}$ gives
\begin{eqnarray}
\langle f \rangle &=& -\frac{1}{c}\langle\textbf{j}\cdot\delta\textbf{A}\rangle + \beta_1 \langle |\phi_0|^4 \rangle (C^4_{\uparrow\!\uparrow} + C^4_{\downarrow\!\downarrow})/2\nonumber\\
&+& \left[\frac{2\gamma}{l^2}\!(\!\langle\!|\phi_0|^2\!|\tilde{\phi}_0|^2\!\rangle\!-\!\langle\!|\phi_0|^2\!|\tilde{\phi}_1|^2\!\rangle\!)
- \beta_2\!\langle\!|\phi_0|^2\!|\tilde{\phi}_0|^2\!\rangle\right]C^2_{\uparrow\!\uparrow} C^2_{\downarrow\!\downarrow}\nonumber\\
&+& \frac{\langle h^2_s \rangle}{8\pi} -
\frac{H^2}{8\pi}+\alpha\left[1-\frac{K+H_{c2}l^2\tilde{\kappa}}{K-H_{c2}l^2\tilde{\kappa}}\right]C_{\downarrow\!\downarrow}^2,
\label{EQ:LLLfree}
\end{eqnarray}
where $\bf{j}$ is the supercurrent, $\delta\bf{A}$ is deviation of
the vector potential from what we would have for the $h = H_{c2}$,
and $h_s$ is the screening field of superconductor. Note that in
this approximation $\bf{j}$ is calculated solely from quadratic
terms, ignoring $\gamma$ terms, and by the Maxwell equation
$\nabla \times \bm{h_s} = 4\pi \bm{j} /c$. More specifically
\begin{eqnarray}
\bm{j}=&2eK[\Delta^*_{\uparrow\!\uparrow}(\bm{D}\Delta_{\uparrow\!\uparrow})+
\Delta^*_{\downarrow\!\downarrow}(\bm{D}\Delta_{\downarrow\!\downarrow})+c.c]\nonumber\\
&-c\tilde{\kappa}\nabla\times\hat{z}(|\Delta_{\uparrow\!\uparrow}|^2-|\Delta_{\downarrow\!\downarrow}|^2)
\end{eqnarray}
Since $\nabla \times \delta \bm{A} = \bm{\hat{z}} (h_s + H - H_{c2})$
and, from Maxwell's equations,
$\nabla \times \bm{h_s} = 4\pi \bm{j} /c$, partial integration leads
to \cite{deGenneSC}
\begin{eqnarray}
\frac{1}{c}\langle\textbf{j}\cdot\delta\textbf{A}\rangle &=& \frac{1}{4\pi}
\langle \bm{h_s}\cdot (\bm{h_s} + \bm{H} - \bm{\hat{z}} H_{c2}) \rangle\nonumber\\
&=& \frac{\langle h^2_s \rangle}{4\pi} + \frac{H_{c2} - H}{4\pi}\langle h_s \rangle.
\end{eqnarray}
Meanwhile, when we calculate the expectation value of $\gamma$, we set
$h = H_{c2}$. This leads to the free energy of
\begin{eqnarray}
\langle f \rangle &=& -\frac{H_{c2}\!-\!H}{4\pi}\langle h_s\rangle - \frac{\langle h^2_s \rangle}{8\pi} - \frac{H^2}{8\pi}\nonumber\\
&+& \alpha\left[1-\frac{K+H_{c2}l^2\tilde{\kappa}}{K-H_{c2}l^2\tilde{\kappa}}\right]C_{\downarrow\!\downarrow}^2 
+ \frac{\beta}{2} \langle |\phi_0|^4 \rangle (C^4_{\uparrow\!\uparrow} + C^4_{\downarrow\!\downarrow})\nonumber\\
&+& \left[\frac{2\gamma}{l^2}\!(\!\langle\!|\phi_0|^2\!|\tilde{\phi}_0|^2\!\rangle\!-\!\langle\!|\phi_0|^2\!|\tilde{\phi}_1|^2\!\rangle\!)
- \beta_2\!\langle\!|\phi_0|^2\!|\tilde{\phi}_0|^2\!\rangle\right]C^2_{\uparrow\!\uparrow} C^2_{\downarrow\!\downarrow}\nonumber\\
&\equiv& - \frac{H^2}{8\pi} +\langle \tilde{f} \rangle.
\label{EQ:freeExpect}
\end{eqnarray}

The screening field $h_s$ can be calculated by assuming
$|\hat{\Delta}| \propto (H_{c2} - H)^{1/2}$ near the second order 
phase transition at $H = H_{c2}$. Here we will deal only with 
$O(|\hat{\Delta}|^4)$ (or equivalently, $O(1-H/H_{c2})^2$) and 
ignore higher order terms. This allows us to calculate $\bf{j}$, 
and consequently $h_s$, solely from quadratic terms. In the 
lowest Landau level, this yields  the screening field:
\begin{eqnarray}
h_s &=& \frac{8\pi^2 K}{\Phi_0} (|\Delta_{\uparrow\!\uparrow}|^2 + |\Delta_{\downarrow\!\downarrow}|^2)
-4\pi \tilde{\kappa}(|\Delta_{\uparrow\!\uparrow}|^2 -|\Delta_{\downarrow\!\downarrow}|^2)\nonumber\\
&=&\!\left(\!\frac{8\pi^2 K}{\Phi_0}\!-\!4\pi\tilde{\kappa}\!\right)\!C^2_{\uparrow\!\uparrow}\!|\phi_0|^2\!
  +\!\left(\!\frac{8\pi^2 K}{\Phi_0}\!+\!4\pi\tilde{\kappa}\!\right)\!C^2_{\downarrow\!\downarrow}\!|\tilde{\phi}_0|^2.\nonumber\\
\label{EQ:screenField}
\end{eqnarray}

Inserting Eq.~(\ref{EQ:screenField}) into Eq.~(\ref{EQ:freeExpect}), 
the free energy takes the following form:
\begin{equation}
\langle \tilde{f} \rangle = -\tilde{\alpha_1} C_{\uparrow\!\uparrow}^2 -\tilde{\alpha_2} C_{\downarrow\!\downarrow}^2
+\tilde{\beta_1} C_{\uparrow\!\uparrow}^4 +\tilde{\beta_2} C_{\downarrow\!\downarrow}^4 + \tilde{\beta_3} C_{\uparrow\!\uparrow}^2 C_{\downarrow\!\downarrow}^2
\label{eq:simplef}
\end{equation}
with terms quadratic or quartic in  $C_{\uparrow\!\uparrow}$ and 
$C_{\downarrow\!\downarrow}$ with coefficients that are independent 
in general (we will further specify these coefficients is the next 
two subsections). In the absence of screening fields, Zeeman-fields 
and the term proportional to $\gamma$, the form of the free energy 
in Eq.~\eqref{EQ:freeExpect} is similar to that examined in 
Ref.~\onlinecite{PhysRevLett.88.180403} in the context of 
two-component Bose condensates (spin-half spinor condensate).In that 
case, the vortex lattice structure is solely determined by the 
competition between the $\beta_1$ and $\beta_2$ terms of 
Eq.~(\ref{EQ:freeLLL1}); the $\beta_1$ term determines the
interaction energy within each vortex lattices, and the $\beta_1$
term the interaction energy between two fractional vortex species
each forming lattices. Specifically, the quartic term
$-\beta_2\langle |\phi_0|^2|\tilde{\phi}|^2\rangle C_{\uparrow\uparrow}^2C_{\downarrow\downarrow}^2$ 
determines the stability of the HQV lattice. If $\beta_2<0$, then 
a HQV lattice is the ground state. If $\beta_2>0$, then full 
quantum vortex lattice is the ground state. In 
Appendix~\ref{app:weakcoupling}, we show that $\beta_2=0$ in 
weak-coupling theories, so that the two lattice structures are 
degenerate. In the rest of the paper, we will focus on aspects that 
are unique to triplet superconductors in subsections 
\ref{subsec:screening} and \ref{subsec:zeeman}: the effects of the 
screening fields $f_{in}^{(4)}$, and the Zeeman-field. 

\subsection{The effects of screening and $f_{in}^{(4)}$}
\label{subsec:screening}

\begin{figure}
\includegraphics[width=0.48\textwidth]{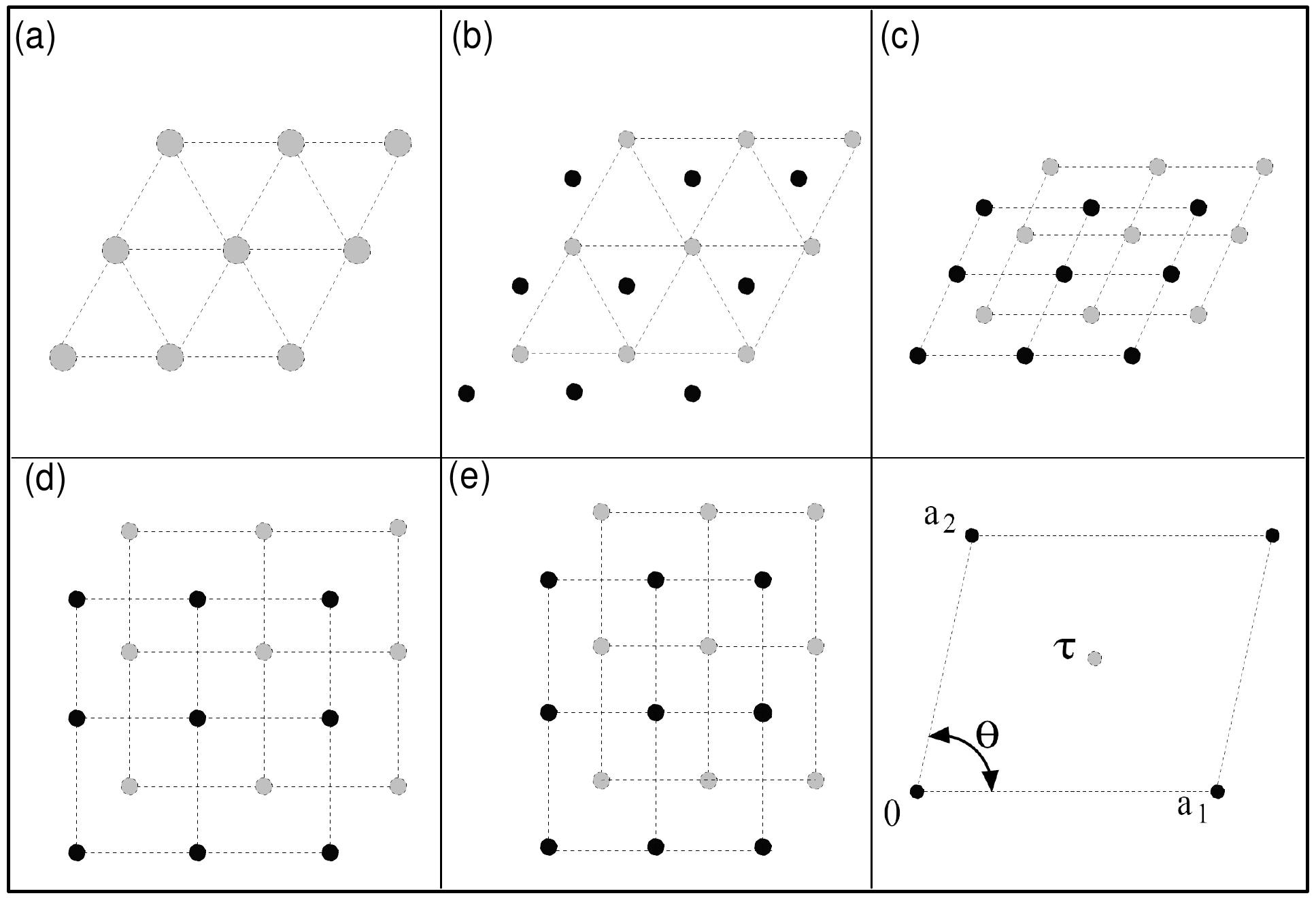}
 \caption{Real space vortex lattice structure}
\label{fig:1a}
\end{figure}

Here we look into the effect of screening. Ignoring the Zeeman 
field (in weak-coupling theories, the Zeeman field is vanishing 
in the clean limit), we have symmetry between 
$C_{\uparrow\!\uparrow}$ and $C_{\downarrow\!\downarrow}$, 
giving us
\begin{equation}
\tilde{\alpha_1} = \tilde{\alpha_2} \equiv
\tilde{\alpha} = \frac{2\pi K(H_{c2}-H)}{\Phi_0}\langle |\phi_0|^2 \rangle
\label{EQ:quadCoeff}
\end{equation}
and
\begin{equation}
\tilde{\beta}_1=\tilde{\beta}_2 \equiv
\tilde{\beta} = \left(\frac{\beta}{2} - \frac{8\pi^3 K^2}{\Phi_0^2}\right)\langle |\phi_0|^4 \rangle,
\label{EQ:quartCoeff}
\end{equation}
which means that the free energy in Eq.\eqref{eq:simplef}
takes a simpler form:
\begin{equation}
\langle \tilde{f} \rangle = -\tilde{\alpha} (C^2_{\uparrow\!\uparrow} + C^2_{\downarrow\!\downarrow}) +
\tilde{\beta}(C^4_{\uparrow\!\uparrow} + C^4_{\downarrow\!\downarrow}) + \tilde{\beta_3} C^2_{\uparrow\!\uparrow} C^2_{\downarrow\!\downarrow},
\label{EQ:freeScreen}
\end{equation}
with
\begin{equation}
\tilde{\beta_3} = \frac{2\gamma}{l^2} \langle |\phi_0|^2 (|\tilde{\phi}_0|^2- |\tilde{\phi}_1|^2)\rangle
- \left(\frac{16\pi^3 K^2}{\Phi_0^2} + \beta_2 \right) \langle |\phi_0|^2 |\tilde{\phi}_0|^2 \rangle.
\label{EQ:quartCoeff2}
\end{equation}
Now the free energy of  Eq.\eqref{EQ:freeExpect}  can be minimized by choosing
$C^2_{\uparrow\!\uparrow} = C^2_{\downarrow\!\downarrow} = \tilde{\alpha}/(2\tilde{\beta} + \tilde{\beta_3})$
(which gives $|\hat{\Delta}| \propto (H_{c2} - H)^{1/2}$ as mentioned), giving us the
free energy expectation value
\begin{equation}
\langle f \rangle = -\frac{H^2}{8\pi} -
\frac{\tilde{\alpha}^2}{2\tilde{\beta}+\tilde{\beta_3}}.
\label{EQ:freeFormulaLLL}
\end{equation}
Eqs.\eqref{EQ:quartCoeff2} and \eqref{EQ:freeFormulaLLL} allows for
understanding the role of both screening and $f_{in}^{(4)}$. The
screening affect the vortex lattice structure through the
dependence of terms proportional to $K^2$ in
Eqs.~(\ref{EQ:quartCoeff}) and (\ref{EQ:quartCoeff2}) on the
lattice structure parameters $\varsigma$, $\sigma$ and ${\bm
\tau}$. Since all quartic expectation values depend on the lattice
structure, the lattice structure will be determined through
minimizing $(2\tilde{\beta}+\tilde{\beta_3})$. Since the magnitude
of $K^2$ term in Eq.\eqref{EQ:quartCoeff2} is larger for full
quantum vortices (Eq.\eqref{EQ:quartCoeff} is not affected),
screening tend to disfavor HQV lattices,  in line with earlier
observation for isolated HQV's\cite{chung:197002}. However, since
weak-coupling theories lie near the point $\tilde{\beta_3}=0$ we
expect the screening effect will put the physical system at a fine
balance between interlacing lattices of HQV's and the ordinary
Abrikosov vortex lattice. Hence it should be possible to observe
the transition between the two phases upon small change of field
and temperature. Indeed, the interaction $f_{in}^{(4)}$ plays this
role. In particular, we numerically find that for $\gamma>0$, then
this term tends to favor HQV lattices. A positive sign of $\gamma$
occurs when $\rho_{sp},\rho_s$ and this is to be expected in
spin-triplet superconductors \cite{chung:197002}. Note, that
unlike the other contributions in $\tilde{\beta}_3$, the
contribution from $f_{in}^{(4)}$ vanishes as $T \rightarrow T_c$.
Consequently, this term can drive a field and temperature
dependent transition between a HQV lattice and a full quantum
lattice. Figures 1 and 2 show the phase diagram for $\gamma=0$.
Note the similarity between the calculated phase diagram and that
found in the context of two-component Bose condensates
\cite{PhysRevLett.88.180403, barnett:240405}.

\begin{figure}
\includegraphics[width=0.45\textwidth]{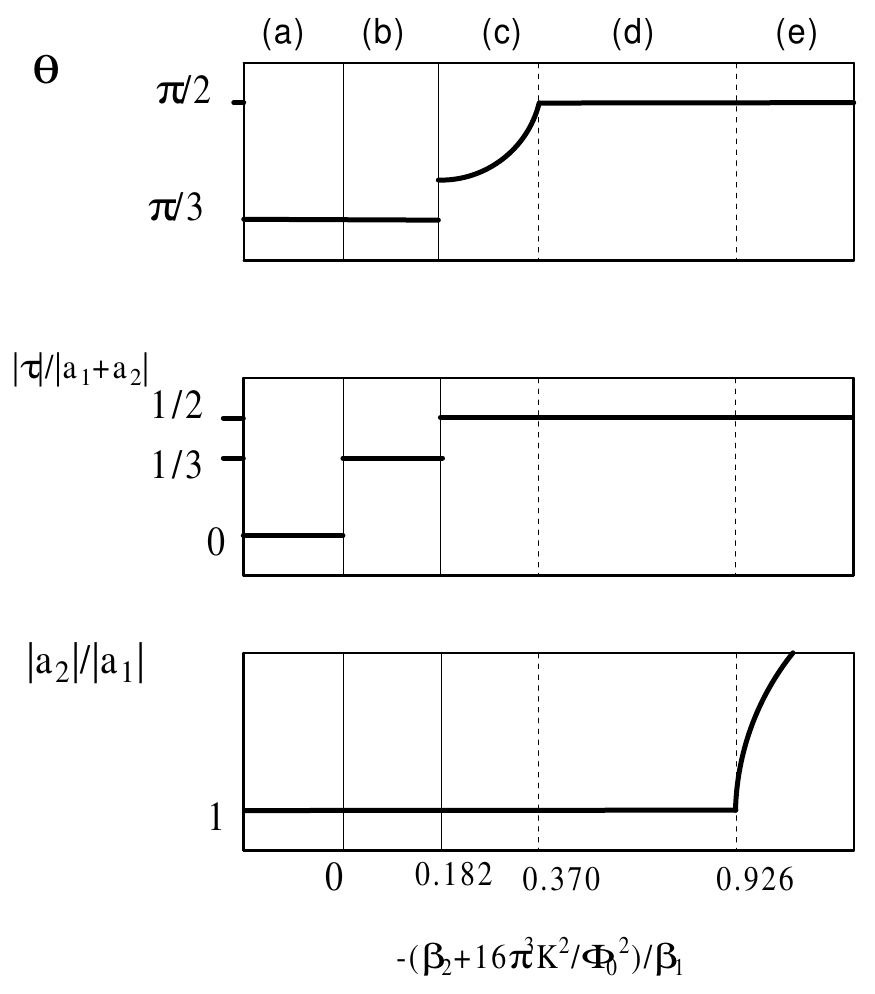}
 \caption{Transition between different vortex lattice structures when $\gamma=0$. (a)-(e) label the vortex lattice structure shown in Fig.~(\ref{fig:1a}).}
\label{fig:1b}
\end{figure}

\subsection{The effects of the Zeeman term}
\label{subsec:zeeman} For simplicity, we consider here the effect
of Zeeman field alone ignoring screening and setting $\gamma=0$
(ignoring $f_{in}^{(4)}$). In the presence of Zeeman field, the
free energy Eq.~\eqref{eq:simplef} would have different
coefficients for two quartic terms:
\begin{equation}
\langle \tilde{f} \rangle = -\tilde{\alpha_1}
C^2_{\uparrow\!\uparrow} -\tilde{\alpha_2}
C^2_{\downarrow\!\downarrow} +
\tilde{\beta}(C_{\uparrow\!\uparrow}^4 +
C_{\downarrow\!\downarrow}^4) + \tilde{\beta_3}
C_{\uparrow\!\uparrow}^2 C_{\downarrow\!\downarrow}^2
\label{eq-free-1}
\end{equation}
where $\tilde{\alpha}_1=\alpha+\frac{K}{l^2}+H\tilde{\kappa}$,
$\tilde{\alpha}_2=\alpha+\frac{K}{l^2}-H\tilde{\kappa}$,
$\tilde{\beta}=\beta_1\langle |\phi_0|^4\rangle$, and
$\tilde{\beta}_3=-\beta_2\langle|\phi_0|^2|\tilde{\phi}_0|^2|^2\rangle$.
The Zeeman field has two main consequences: (i) it typically 
leads to two phase transitions. In the first phase 
$C_{\uparrow\!\uparrow}\ne 0$ and $C_{\downarrow\!\downarrow} = 0$
and in the second phase, both components are non-zero. The first
phase is analogous to the $^3$He A$_1$ phase, with a non-unitary
spin-triplet order parameter. However, weak-coupling theories
prefer unitary spin-triplet states and this drives the second 
transition. (ii) In the fractional vortex lattice phase where 
both components are non-zero, the magnetic flux contained by 
isolated fractional vortices is no longer a half-integral flux 
quanta. Instead, two types of vortices each carry fractional 
flux values of 
\begin{equation}
\Phi_i=\Phi_0\frac{|c_i|^2}{|c_1|^2+|c_2|^2}.
\end{equation}

The double transition is possible if 
$2\tilde{\beta}+\tilde{\beta_3}>0$ (note again that weak-coupling 
theories yield $\tilde{\beta_3}=0$ and $\tilde{\beta}>0$), there 
can be two transitions with a second transition appearing at a 
temperature
\begin{equation}
T_{c2}-T_{c1}=\frac{4\tilde{\beta}}{2\tilde{\beta}+\tilde{\beta}_3}\frac{K+H_{c2}l^2\tilde{\kappa}}{K_1-H_{c2}l^2\tilde{\kappa}}T_{c1}.
\end{equation}
In the high temperature phase, the vortex lattice is hexagonal
and, at the second transition, the lattice will remain hexagonal
and the second component will either coincide with first or be
displaced half a hexagonal vortex lattice vector from the first.
As temperature is further reduced below the second transition, the
lattice will continuously deform, asymptotically approaching the
phases presented in the subsection ~\ref{subsec:screening}
(those shown in Fig.~2). The resulting phase diagram is
qualitatively shown in Fig.~3.

%Note that the quadratic coefficients are no longer equal. {\it This means that
Both consequences of the Zeeman field stem from breaking the
additional $\mathbb{Z}_2$ symmetry that is present when
$\tilde{\alpha}_1=\tilde{\alpha}_2$. In general, the existence of
fractional vortices is the result of the $U(1)\times U(1)$
symmetry of the free energy. When there is additional 
$\mathbb{Z}_2$ symmetry due to $\tilde{\alpha}_1=\tilde{\alpha}_2$, 
the flux contained in each fractional vortices are restricted to be 
half the flux quantum since the two components of the order 
parameter are no longer degenerate in a magnetic field. In Section 
VII, we will see that this helps us distinguish a lattice of HQVs 
from a lattice of full quantum vortex.

\begin{figure}
\includegraphics[width=0.45\textwidth]{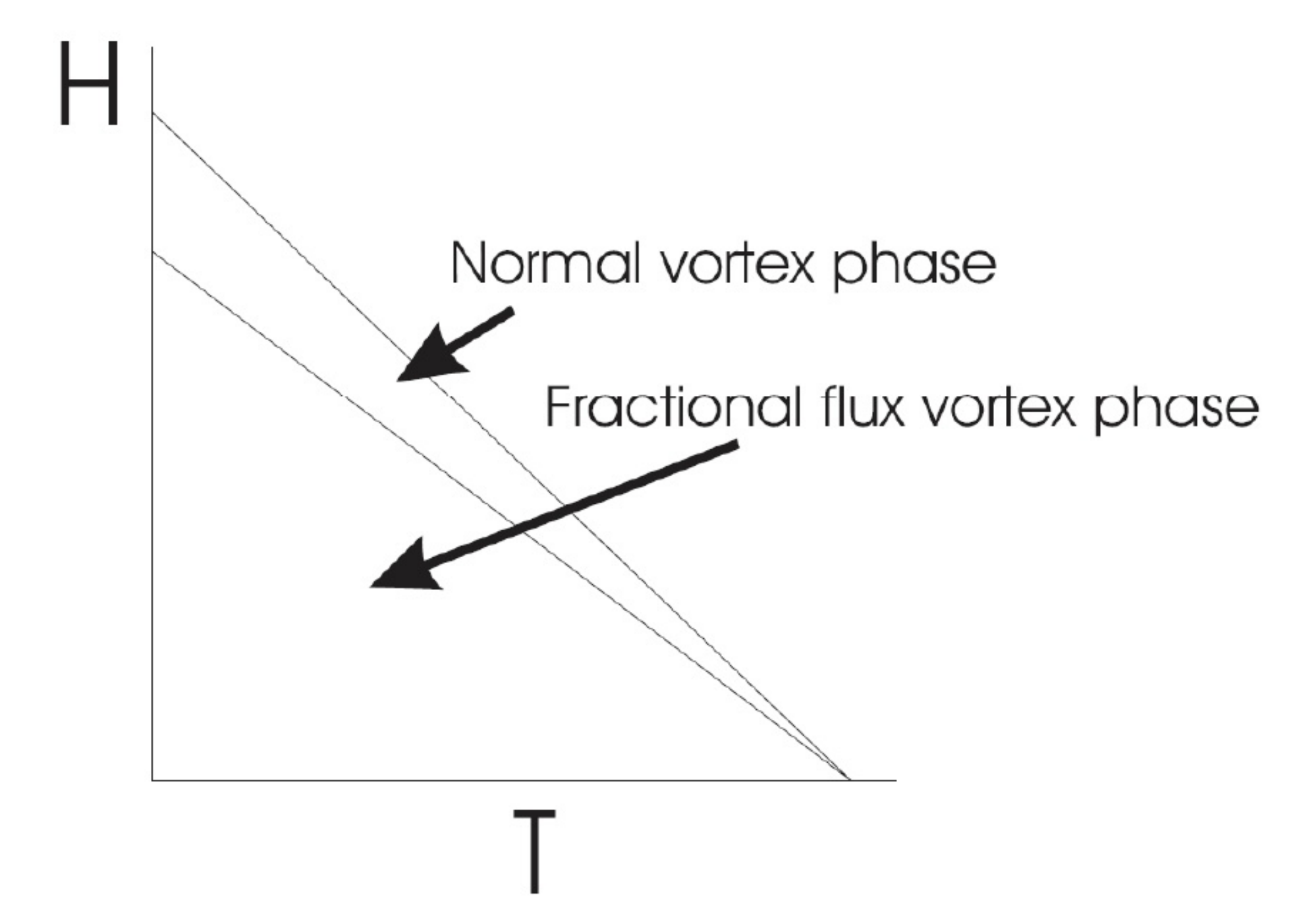}
 \caption{The role of the Zeeman field is to cause two transitions 
and to change the HQV lattice phase to a a phase with two types of 
fractional vortices in which the two fractional fluxes sum to 
$\Phi_0$. Far below the second transition it is expected that these 
fractions will be well approximated by $1/2$.}
\label{fig2}
\end{figure}

\subsection{Chiral Triplet superconductors: lowest Landau level solution}
\label{subsec:chiral}
The chiral triplet superconductor with tetragonal symmetry, because
of the inhomogeneous quadratic terms we have already seen in
Eq.~(\ref{EQ:quadDeriv2}),
\begin{eqnarray}
f^{(2,s)}_{in} &=& K_1(|{\bf D}\Delta_{s,+}|^2 + |{\bf D}\Delta_{s,-}|^2)\nonumber\\
&+&K_2[ \!\{\!(D_x \Delta_{s,+})\!(D_x \Delta_{s,-})^*\!-\!(D_y \Delta_{s,+})\!(D_y \Delta_{s,-})^*\!\}\!/2\nonumber\\
&+& \!\{\!(D_x \Delta_{s,-})\!(D_x \Delta_{s,+})^*\!-\!(D_y \Delta_{s,-})\!(D_y \Delta_{s,+})^*\!\}\!/2\nonumber\\
&+& i\!\{\!(D_x \Delta_{s,-})\!(D_y \Delta_{s,+})^*\!+\!(D_y \Delta_{s,-})\!(D_x \Delta_{s,+})^*\!\}\!/2\nonumber\\
&-& i\!\{\!(D_x \Delta_{s,+})\!(D_y \Delta_{s,-})^*\!+\!(D_y \Delta_{s,+})\!(D_x \Delta_{s,-})^*\!\}\!/2]\nonumber\\
&+& K_4 (|D_z\Delta_{s,+}|^2 + |D_z\Delta_{s,-}|^2),
\end{eqnarray}
has a much more complicated quadratic free energy,
\begin{equation}
f^{(2)}_0 = \sum_{s=\uparrow\!\uparrow,\downarrow\!\downarrow}[-\alpha(|\Delta_{s,+}|^2+|\Delta_{s,-}|^2)+f^{(2,s)}_{in}],
\label{EQ:quadChiral0}
\end{equation}
even when we exclude the Zeeman field and any spin-orbit coupling.

Due to these inhomogenous quadratic terms, we cannot put both
chirality components in the lowest Landau level. This is due to the
presence of
$[(D_x \Delta_{s,\sigma})(D_y \Delta_{s,-\sigma})^* + {\rm c.c}]$
terms in $f^{(2,s)}_{in}$. The above quadratic free energy of
Eq.~(\ref{EQ:quadChiral0}), together with Eq.~(\ref{EQ:chiralSplit})
that gives us the energy splitting between two chiralities, leads to
the linearized GL equation
\begin{eqnarray}
&\,&\alpha l^2\!\left(\!\begin{array}{c}
\Delta_{s+} \\ \Delta_{s-}\\
\end{array}\!\right) =\nonumber\\
&\,&\left[\!\begin{array}{cc}
K_1\!(\!1\!+\!2\Pi_+\Pi_-\!)\!-\!\delta K & K_2\Pi_-^2\\
K_2\Pi_+^2 & K_1\!(\!1\!+\!2\Pi_+\Pi_-\!)\!+\!\delta K\\ \end{array}\!\right]\!\left(\!\begin{array}{c}
\Delta_{s+} \\ \Delta_{s-}\\
\end{array}\!\right).\nonumber\\
\label{EQ:matrixGL}
\end{eqnarray}

However, a lowest Landau level solution that satisfies
Eq.~(\ref{EQ:matrixGL}) may still have the highest $H_{c2}$ and
therefore be possible \cite{Zhitomirskii_UPt3}. This would lead to
a nonzero order parameter for only one chirality -
$(\Delta_{s+},\Delta_{s-}) = C(0,\phi_0)$ - and requires $\delta K
< -\frac{K_2^2}{4K_1}$. In this case, the vortex energetics of the
chiral triplet superconductor is identical to that of the
nonchiral triplet superconductor, for inserting this lowest Landau
level solution into the full Gibbs free energy leads us back to
Eq.~(\ref{EQ:LLLfree}).

However, Eq.~(\ref{EQ:matrixGL}) can also give us a solution with
Landau level mixing; in this case, both $\Delta_{s+}$ and
$\Delta_{s-}$ are nonzero. We will present discussion on this Landau
level mixing in Appendix~\ref{sec:mixLandau}.

\section{Role of spin-orbit Coupling}
\label{sec:SO}
\subsection{Generic Case}

If the material has orthogonal or tetragonal symmetry (though not
necessarily true for hexagonal symmetry, which is discussed in the
next subsection), then there will exist spin-orbit coupling terms
of the type
\begin{equation} \epsilon \Delta_{\uparrow\uparrow}\
\Delta^*_{\downarrow\downarrow}.
\end{equation}
Such terms break the $U(1)\times U(1)$ symmetry and
consequently isolated fractional flux vortices are no longer
stable. Nevertheless, a fractional flux quantum vortex lattice can
still exist, provided that the separation between vortices is less
than $\xi_{so}$ defined through $\xi_{so}^2=K/\epsilon$.

For completeness, we write here the spin-orbit coupling terms that
appear in the context of a chiral spin-triplet superconductor.
While we do not include these terms in calculations, they may be
useful in other contexts. Due to the tetragonal $C_4$ symmetry,
homogeneous spin-orbit coupling terms should be invariant under
the transformation
\begin{eqnarray}
(\Delta_{\uparrow\!\uparrow,+},\Delta_{\uparrow\!\uparrow,-},\Delta_{\downarrow\!\downarrow,+},\Delta_{\downarrow\!\downarrow,-})
\to  \nonumber \\
(-\Delta_{\uparrow\!\uparrow,+},\Delta_{\uparrow\!\uparrow,-},\Delta_{\downarrow\!\downarrow,+},-\Delta_{\downarrow\!\downarrow,-}).
 \nonumber \end{eqnarray}
To the quadratic order, this condition is satisfied by \footnote{In the absence of out-of-plane magnetic field, such spin-orbit interaction support superconductor with order parameter symmetry analogous to that of $^3$He-B phase. This class of phase has been recently discussed as a time-reversal invariant topological superconductor \cite{topoSC}.}
\begin{eqnarray}
f^{(2)}_{SO} &=& \!\epsilon_1(\!|\Delta_{\uparrow\!\uparrow,+}|^2\!+\!|\Delta_{\downarrow\!\downarrow,-}|^2\!-\!|\Delta_{\uparrow\!\uparrow,-}|^2\!-\! |\Delta_{\downarrow\!\downarrow,+}|^2\!)\nonumber\\
&+& \epsilon_2[(\!\Delta_{\uparrow\!\uparrow,-}\!)\!(\!\Delta_{\downarrow\!\downarrow,+}\!)^* + {\rm c.c.}]\nonumber\\
&+&
\epsilon_3[(\!\Delta_{\uparrow\!\uparrow,+}\!)\!(\!\Delta_{\downarrow\!\downarrow,-}\!)^*
+ {\rm c.c.}].
\end{eqnarray}
We note here that $\epsilon_i$'s can be estimated from the recent ARPES data \cite{haverkort:026406,liu:026408}.

\subsection{Hexagonal Materials}

For hexagonal materials, there exist spin-triplet pairing states
for which no such terms such as that in the above equation appear.
These states belong to the two-dimensional representations
labelled $\Gamma_{5}^-$ and $\Gamma_6^-$ in the review article by
Sigrist and Ueda\cite{RevModPhys.63.239}. Consequently, these materials need
to be considered more carefully. 

We will now show that in hexagonal materials, a little away from $H_{c2}$, 
spin-orbit coupling does not break $U(1) \times U(1)$ symmetry.
For hexagonal materials, the only term that exists in the GL free
energy that is due to spin-orbit coupling is (note that the
inclusion of this term gives rise the complete free energy found
that is found in Sigrist and Ueda for the $\Gamma_{5,6}^-$
representations):
\begin{eqnarray}
f_{SO} &=& K_{so}\Big \{(D_x \Delta_{\downarrow\downarrow})(D_x
\Delta_{\uparrow \uparrow})^*-(D_y \Delta_{\downarrow\downarrow})
(D_y \Delta_{\uparrow\uparrow})^* \nonumber\\
&+&(D_x \Delta_{\uparrow \uparrow})(D_x
\Delta_{\downarrow\downarrow})^*-
(D_y \Delta_{\downarrow\downarrow})(D_y \Delta_{\uparrow\uparrow})^*\nonumber\\
&-& i[(D_x \Delta_{\downarrow\downarrow})(D_y
\Delta_{\uparrow\uparrow})^*+
(D_y \Delta_{\downarrow\downarrow})(D_x \Delta_{\uparrow\uparrow})^*]\nonumber\\
&+& i[(D_x \Delta_{\uparrow\uparrow})(D_y
\Delta_{\downarrow\downarrow})^*+
(D_y \Delta_{\uparrow\uparrow})(D_x \Delta_{\downarrow\downarrow})^*]\Big \}/2\nonumber\\
\end{eqnarray}
With the field along the $c$-axis, the solution to the quadratic
problem satisfies
\begin{equation}
\frac{\alpha l^2}{K}\!\left( \begin{array}{c}
\Delta_{\uparrow \uparrow} \\ \Delta_{\downarrow \downarrow}\\
\end{array} \right)\!=\!
\!\left( \begin{array}{cc}
1+2N-K_z & \tilde{K}_{so}\Pi_-^2\\
\tilde{K}_{so}\Pi_+^2 & 1+2N+K_z\\
\end{array} \right) \left(
\begin{array}{c}
\Delta_{\uparrow \uparrow} \\ \Delta_{\downarrow \downarrow}\\
\end{array} \right)
\nonumber
\end{equation}
where $K_z=\frac{\tilde{K}Hl^2}{K}$ and $\tilde{K}_{so}=K_{so}/K$.
All the eigenstates for this problem can be found analytically\cite{superRev}.
Typically, $|\tilde{K}_{so}|<<1$, so we will be interested in
the eigenstates that contain the lowest Landau level (which will
minimize the free energy when $\tilde{K}_{so}=0$). The two
relevant eigenstates that we wish to keep are:
$(\Delta_{\uparrow \uparrow}, \Delta_{\downarrow,\downarrow})=(\phi_0,\epsilon \phi_2)$
and $(\Delta_{\uparrow \uparrow},\Delta_{\downarrow,\downarrow})=(0,\phi_0)$
where $\epsilon$ is proportional to $\tilde{K}_{so}$. 
Note that unlike in the subsection~\ref{subsec:chiral} we can keep 
both solutions because while we examined $H \sim H_{c2}$ in that subsection, 
we are a little away from $H_{c2}$ in this subsection.
We therefore write $\Delta_{\uparrow\uparrow}=\gamma_1 \phi_0$ and
$\Delta_{\downarrow\downarrow}=\gamma_1 \epsilon\phi_2+\gamma_2\tilde{\phi}_0$
to include these two eigenstates. For simplicity, we ignore
screening and the Zeeman field to find the following free energy
\begin{eqnarray}
\langle f \rangle =&&-(1-H/H_{c2,1})|\gamma_1|^2-(1-H/H_{c2,2})|\gamma_2|^2\nonumber\\
&&+\beta_1[|\gamma_1|^4\langle |\phi_0|^4\rangle+\langle|\gamma_1\epsilon\phi_2+\gamma_2\tilde{\phi}_0|^4\rangle]\nonumber\\
&&-\beta_2\langle |\gamma_1\phi_0|^2|\gamma_1\epsilon\phi_2+\gamma_2\tilde{\phi}_0|^2\rangle
\end{eqnarray}
where $H_{c2,i}$ ($i=1,2$) is the upper critical field for
eigenstate $i$. Since spin-orbit coupling is expected to be small,
this implies that $\epsilon<<1$, so keeping to linear order in
$\epsilon$ yields:
\begin{eqnarray}
\langle f \rangle =&&-(1-H/H_{c2,1})|\gamma_1|^2-(1-H/H_{c2,2})|\gamma_2|^2\nonumber\\
&&+\beta_1\langle|\phi_0|^4\rangle(|\gamma_1|^4+|\gamma_2|^4)+\beta_2\langle|\phi_0|^2|\tilde{\phi}_0|^2\rangle|\gamma_1|^2|\gamma_2|^2\nonumber\\
&&+\epsilon[\beta_1|\gamma_2|^2\gamma_2\gamma_1^*\langle|\tilde{\phi}_0|^2\tilde{\phi}_0\phi_2^*\rangle+{\rm c.c}]\nonumber\\
&&-\epsilon[\beta_2|\gamma_1|^2\gamma_1\gamma_2^*\langle|\phi_0|^2\phi_2\tilde{\phi}_0^*\rangle + {\rm c.c}]
\label{eq-so}
\end{eqnarray}
Without the last two terms, this theory is the same as that found
for non-chiral spin-triplet superconductors with a Zeeman field but
without any spin-orbit coupling. At the upper critical field, one of
the two components $\gamma_1$ or $\gamma_2$ order and the vortex
lattice will be hexagonal (this conclusion is correct even when
including terms that are second order in $\epsilon$). As the
temperature or magnetic field is reduced, the last two terms in
Eq.~(\ref{eq-so}) can play an important role. These two terms break
the $U(1)\times U(1)$ symmetry of the theory and therefore will
tend to remove any HQV lattice phases. However, the spatial averages
$\langle|\tilde{\phi}_0|^2\tilde{\phi}_0\phi_2^*\rangle$ and
$\langle|\phi_0|^2\phi_2\tilde{\phi}_0^*\rangle$ {\it vanish} for
a hexagonal vortex lattice (loosely speaking, this follows from
noting that $\phi_n$ picks up a factor $e^{in\phi}$ under a
rotation about $\hat{z}$ and that a hexagonal vortex lattice is
symmetric under rotations of $\pi/3$) . The hexagonal symmetry of
the materials conspires to remove this form of $U(1)\times U(1)$
symmetry breaking and the HQV lattice structures are still
possible (indeed the theory is the same as that given for the
non-chiral spin-triplet superconductors with a Zeeman field, but without
spin-orbit coupling). Note that if ${\bm \tau}\ne 0$ (signaling the existence of the
fractional vortex lattice), then
$\langle|\tilde{\phi}_0|^2\tilde{\phi}_0\phi_2^*\rangle$=
$\langle|\phi_0|^2\phi_2\tilde{\phi}_0^*\rangle$=0 for any lattice
geometry. Consequently, the last two terms of Eq.~\eqref{eq-so}) do
not play any role in the theory of the fractional vortex lattices.

It is reasonable to ask if there are any other $U(1)\times U(1)$
symmetry breaking terms that we have neglected in the above
analysis. Indeed there is one that appears at order $\epsilon^2$:
$\epsilon^2\beta_1\gamma_1^2(\gamma_2^2)^*\langle\phi_2^2(\tilde{\phi}_0^2)^*\rangle$.
This term allows for the existence of a fractional vortex lattice
phase subject to the constraint that ${\bm \tau}$ is half a vortex
lattice translation vector\cite{superRev}.
There are also $U(1)\times U(1)$ that appear at order
$\epsilon^3$, but these vanish for the same reason as the order
$\epsilon$ term. Consequently, the spin-orbit coupling for the
hexagonal two-dimensional representations plays essentially the
same role as the Zeeman field.

\section{Observation of the Vortex Lattice}
\label{sec:observation}

\begin{figure*}[t]
%  \subfloat[]{
   \includegraphics[width=.27\textwidth]{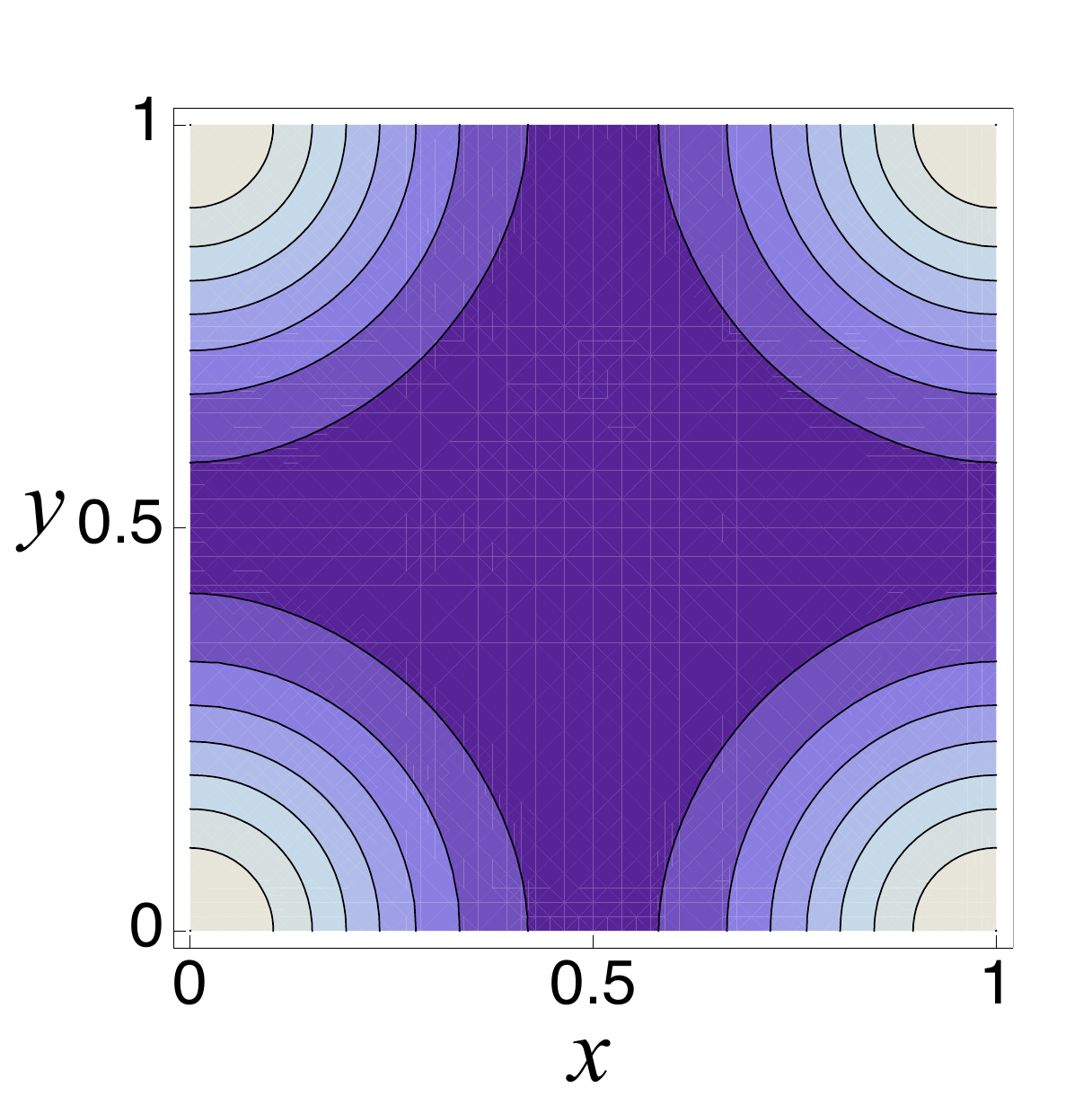}
%   }
%    \subfloat[]{
   \includegraphics[width=.27\textwidth]{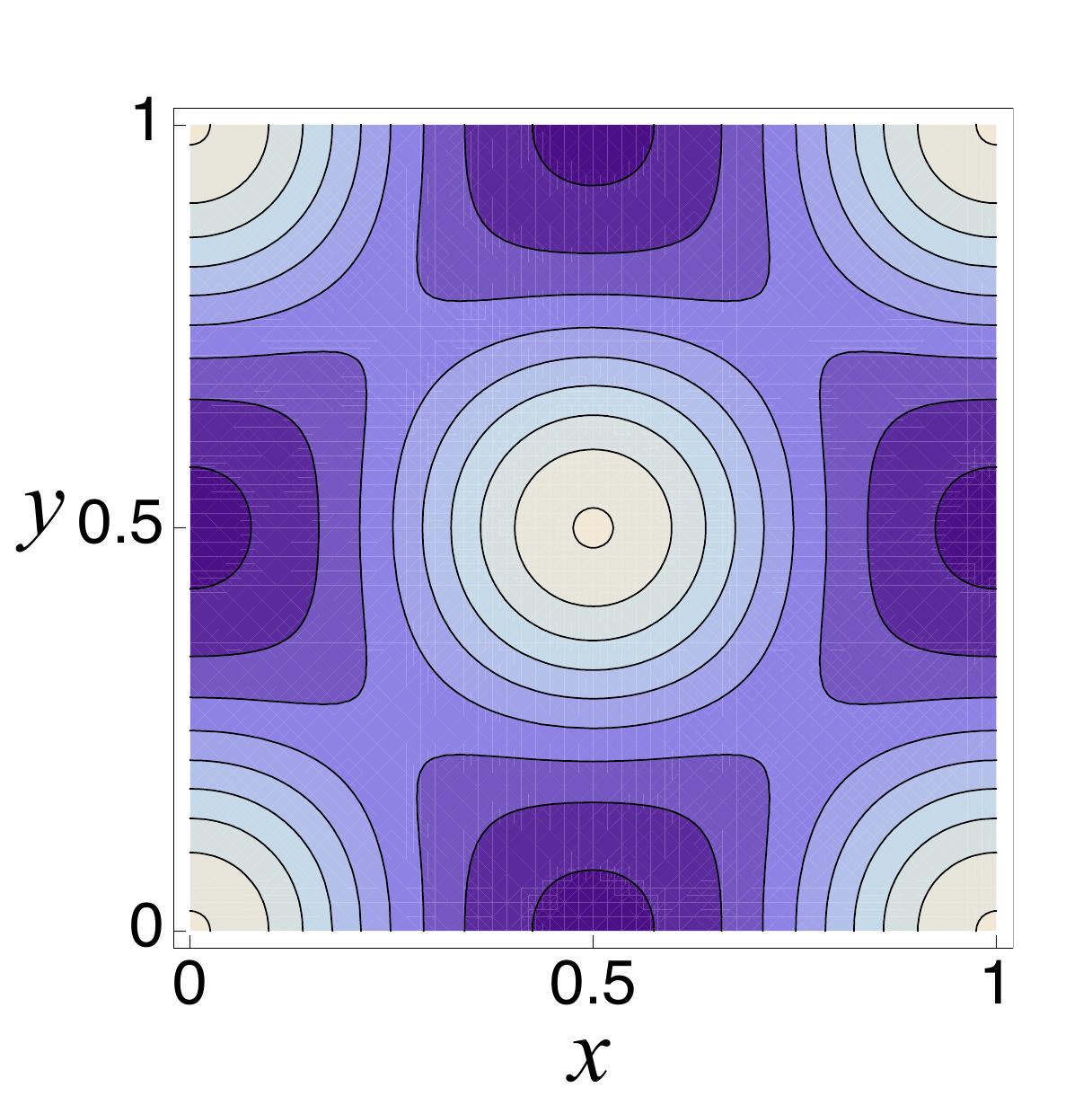}
%   }
%\subfloat[]{
   \includegraphics[width=.27\textwidth]{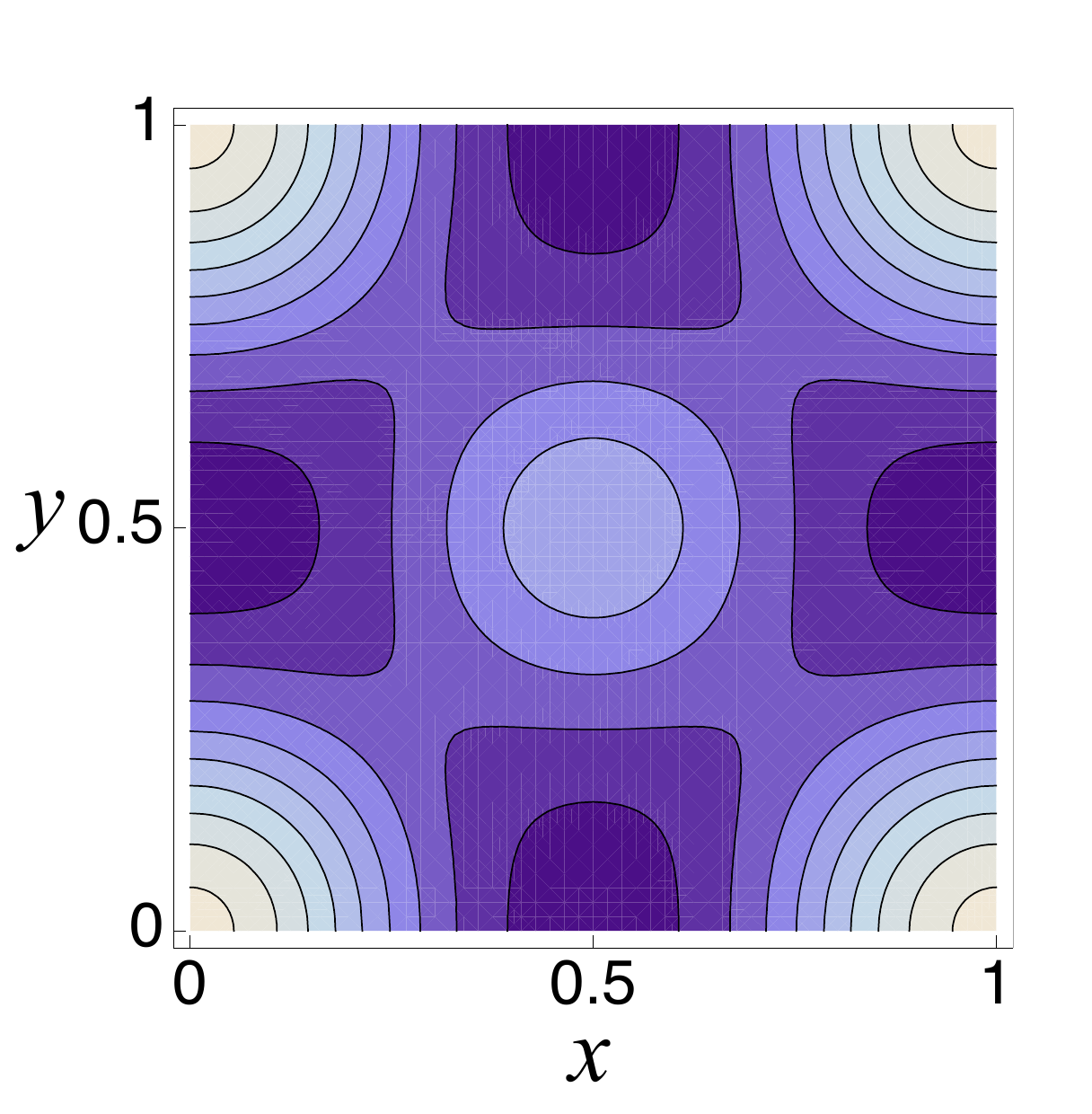}
%   }
   \caption{The contour plots of the screening field in the real (that is, position) space for different types of vortex lattices: (a) a lattice of ordinary Abrokosov (full quantum) vortices (b) a HQV lattice (c) a lattice of fractional vortices in the presence of Zeeman field.
 The position is measured in units of magnetic length.
   Note the halving of the unit cell in going from the HQV lattice to the full quantum vortex lattice.
   When the Zeeman field is added in, the periodicity is that of the full quantum lattice, though the vortex lattice unit cell
 now has additional structure due to appearance of fractional flux at both the corners and center of the unit cell.}
   \label{fig:structure}
\end{figure*}

The best way to determine both the vortex lattice structure and
the vortex type is to observe the magnetic field distribution through
the small angle neutron scattering. What we will see in this experiment
is the Fourier transform $f({\bf G})$ of the screening field of
Eq.~(\ref{EQ:screenField}),
\begin{eqnarray}
h_s({\bf r}) &=& (\frac{8\pi^2 K}{\Phi_0}-4\pi\tilde{\kappa}) C^2_{\uparrow\!\uparrow} |\phi_0({\bf r})|^2\nonumber\\
&+& (\frac{8\pi^2 K}{\Phi_0}+4\pi\tilde{\kappa})C^2_{\downarrow\!\downarrow} |\tilde{\phi}_0({\bf r})|^2
\end{eqnarray}
- that is $h_s({\bf r}) = \sum_{\bf G} f({\bf G}) \exp(i {\bf G} \cdot {\bf r})$,
where ${\bf G}$ is the reciprocal lattice vectors of the vortex lattice
in the unit of the inverse magnetic length.

The characteristic feature of the vortex lattice with half-vortices
in the small angle neutron scattering experiment is the modulation
of the Bragg peaks. The form factor of the Bragg peaks would be
the $f({\bf G})$ of the last paragraph. Using
\begin{equation}
|\phi_0({\bf r})|^2 = \sum_{\bf G} (-1)^{m_1 + m_2 + m_1 m_2} e^{-{\bf G}^2/2}
\end{equation}
where ${\bf G} = m_1 {\bf G_1} + m_2 {\bf G_2}$, and ${\bf G_i}$'s are the basis
vector of the reciprocal lattice, we obtain the form factor
\begin{eqnarray}
f({\bm G}) &=& (-1)^{m_1 + m_2 + m_1 m_2} e^{-{\bm G}^2/2}[(\frac{8\pi^2 K}{\Phi_0}-4\pi\tilde{\kappa})|C_{\uparrow\!\uparrow}|^2\nonumber\\
&+& (\frac{8\pi^2 K}{\Phi_0}+4\pi\tilde{\kappa})|C_{\downarrow\!\downarrow}|^2 e^{i{\bm G}\cdot {\bm \tau}}].
\end{eqnarray}
This equation implies that the intensity $|f({\bm G})|^2$ for our Bragg
peaks does not come out same for all ${\bf G}$'s. This is because for
almost all vortex lattice structure (the single exception being not very
robuts honeycomb lattice) ${\bm \tau}$
is half a vortex translation vector so that
we have $e^{i{\bm G}\cdot {\bm \tau}}=-1$ for half of ${\bf G}$'s
and $e^{i{\bm G}\cdot {\bm \tau}}=1$ for the other half. When there is
no Zeeman field, $e^{i{\bm G}\cdot {\bm \tau}}=-1$ peaks disappear completely;
natural given that magnetic field cannot distinguish the spin up and the spin
down HQV's at all and thus sees the unit lattice vector halved. However,
when the Zeeman field breaks down the $\mathbb{Z}_2$ symmetry between the
spin up-up pairs and down-down pairs, we now see a secondary peak for
$e^{i{\bm G}\cdot {\bm \tau}}=-1$ as shown on Fig.~(\ref{fig:structure}).

Another promising direction for detecting fractional vortex
lattice would be to use spin-polarized STM to probe the vortex
cores. The key point is that the low energy quasi-particle spins
have opposite polarization in the two different HQV's. This is
because for half of HQV cores, we have
$\Delta_{\uparrow\uparrow}=0$ and
$\Delta_{\downarrow\downarrow}\ne 0$, so that only spin-down
quasi-particles are gapped. On the other hand, for the other half
of HQV cores, only spin-up quasi-particles are gapped. This
spin-polarization of the subgap core modes should be readily
detected through spin-polarized STM.

\section{Conclusion}
\label{sec:conclusion} In this paper we explored various
possibilities for fractional vortex lattice structures in spin
triplet superconductors starting from the most general from of
Gibbs free energy that is allowed by the symmetry of the order
parameter and that of the lattice symmetries relevant for three
candidate spin triplet superconductors, namely  single layer
ruthenate Sr$_2$RuO$4$\cite{Maeno:1994fk,RevModPhys.75.657}
cobaltate Na$_x$CoO$_2 \cdot y$H$_2$O\cite{Nature.Physics.1.91}
and organic\cite{organic} (TMTSF)$_2$ClO$_4$.
 The focus of our analysis was on the role of aspects unique to triplet superconductors, such as
(i) Cooper pair Zeeman field,  (ii) spin-orbit coupling, (iii)
screening, and (iv) interaction effects in the energetics of the
vortex lattice structure.  (i) The Cooper pair Zeeman field breaks
a $\mathbb{Z}_2$ symmetry of the free energy whose presence
constrains the fractional vortices to contain half integral flux
quanta. The resulting structure is that of two interlacing
lattices of vortices containing arbitrary fraction of flux quanta
that adds up to one flux quanta.  Such fractional vortex lattices
will have interesting field distributions in vortex lattice unit
cell due to internal structures within the unit cell. (ii) The
effect of spin-orbit coupling is lattice symmetry specific. In
hexagonal lattices systems such as cobaltates
Na$_x$CoO$_2\cdot$yH$_2$O, spin-orbit coupling has the same effect
as the Cooper pair Zeeman field, supporting fractional vortex
lattices.  However, for tetragonal or orthorhombic lattices,
 sufficiently strong spin-orbit coupling generally favors
 ordinary Abrikosov vortex lattice over HQV's. However,
 such an effect is  relatively 
mild in a dense vortex lattice, provided that the separation between the HQV vortices is less
that a length set by the spin-orbit coupling.
(iii) The Meissner screening effectively generates attraction
between two HQV's with opposite winding of the spin phase and
weakly destablizes the HQV's. (iv) The interaction effects clearly
support energetic stability of HQV's within the GL theory. The
interaction effects  represented by inhomogeneous (unique to
triplet superconductors) quartic terms can drive difference in
effective superfluid stiffness $\rho_{sp} < \rho_s$ which
stabilizes HQV's in the London limit. When the above effects are
put together,  all weak coupling theories we examined appears to
lie at the point of fine balance between ordinary Abrikosov vortex
lattice and lattices of HQV's. Hence it should be possible to
observe transitions between these structures with small changes of
parameters. This further motivates experimental search for these
fractional vortex lattices. We have sketched possible routes for
such searches using neutron scattering or spin polarized STM.

\noindent {\bf Acknowledgements} We are grateful to H.\ Bluhm for
helpful discussions regarding inhomogeneous quartic term and M.\
Sigrist for discussions on spin-orbit coupling. We thank M.\
Stone, D.\ Podolsky, S.\ Mukerjee, E.\ Berg, S.\ Raghu for
numerous useful discussions. E-AK was supported in part by the
Cornell Center for Materials Research (CCMR) through NSF Grant No.
DMR 0520404. SBC was supported by the Stanford Institute of
Theoretical Physics and NSF Grant No. DMR 06-03528. We acknowledge
the KITP for its hospitality through the miniprogram
``Sr$_2$RuO$_4$ and Chiral p-wave Superconductivity" during
initial stages of this work.

\appendix

\section{Ginzburg Landau Energy: Fourth order terms from weak-coupling theory}
\label{app:weakcoupling}

The GL free energy can be determined in the weak-coupling limit.
In the context of the existence of 1/2 qv lattice structures, the
result for the fourth order terms in the free energy turns out to
be highly relevant.  As shown here, this reveals that
weak-coupling theories sit at a point in which the 1/2 qv and the
full qv lattices are degenerate. This indicates that interactions
beyond the weak-coupling limit are essential to determining the
which lattice structure actually appears (screening plays a role
here as well as shown earlier).

The portion of the free energy we calculate here is given in Eq.
3.6
\begin{equation}
f^{(4)}_{hom}=\beta_1(\sum_{i}|\Delta_{i}|^2)^2+\beta_2|\Delta_{\uparrow\!\uparrow}|^2
|\Delta_{\downarrow\!\downarrow}|^2.
\end{equation}
The weak-coupling limit (without spin-orbit coupling) yields (this
follows from Ref.~\onlinecite{RevModPhys.63.239})
\begin{equation}
f^{(4)}_{hom}\propto \langle|{\bf d}({\bf
k})|^4\rangle+\langle{\bf q}^2({\bf k})\rangle
\end{equation}
where ${\bf q}({\bf k})=i{\bf d}({\bf k})\times {\bf d}^*({\bf
k})$, $\langle h({\bf k}) \rangle$ means average $h({\bf k})$ over
all ${\bf k}$ on the Fermi surface, and the proportionality
constant can be found but it is not important for our
considerations. When ${\bf q}$ is non-zero, then the
superconducting state is called non-unitary. In weak-coupling
theories, non-unitary states cost energy and typically do not
appear. Using the gap structure of Eq. 2.2, we find
\begin{eqnarray}
f^{(4)}_{hom}&& \propto \langle|f({\bf
k})|^4\rangle[(|\Delta_{\uparrow\uparrow}|^2+|\Delta_{\downarrow\downarrow}|^2)^2+(|\Delta_{\uparrow\uparrow}|^2-|\Delta_{\downarrow\downarrow}|^2)]
\nonumber\\ && = 2\langle|f({\bf
k})|^4\rangle(|\Delta_{\uparrow\uparrow}|^4+|\Delta_{\downarrow\downarrow}|^4).
\end{eqnarray}
This implies that $\beta_2=-2\beta_1$, independent of the shape of
the Fermi surface. The lack of interaction between the two
components of the gap function leads to the degeneracy between the
1/2 qv and the full qv lattice structures.

\section{Ruthenate - the Landau level mixing}
\label{sec:mixLandau}

We show here how we can have the Landau level mixing in a chiral
triplet superconductor. The case we are considering here is in
the weak pairing regime and has tetragonal crystalline symmetry
and a cylindrical Fermi surface. Let us consider again the linearized
GL equation:
\begin{equation}
l^2\!\left( \begin{array}{c}
\Delta_{s+} \\ \Delta_{s-}\\
\end{array} \right)\!=\!
\frac{K}{\alpha}\!\left( \begin{array}{cc}
1+2\Pi_+\Pi_- & \Pi_-^2\\
\Pi_+^2 & 1+2\Pi_+\Pi_-\\ \end{array} \right) \left(
\begin{array}{c}
\Delta_{s+} \\ \Delta_{s-}\\
\end{array} \right),
\end{equation}
where $s = \uparrow\!\uparrow, \downarrow\!\downarrow$.
(Note that, though otherwise same as Eq.~(\ref{EQ:matrixGL}),
we now ignore the energy splitting between the $\pm$
chiralities and set $K_1 = K_2 = K$.) This matrix equation
as a solution in the form
\begin{eqnarray}
\left(
\begin{array}{c}
\Delta_{s+} \\ \Delta_{s-}\\
\end{array} \right) = C_s \left(
\begin{array}{c}
\phi_0 \\ -\delta \phi_2 \\
\end{array} \right),
\label{EQ:landauMixingSol}
\end{eqnarray}
where $\delta = \sqrt{3} - \sqrt{2}$.

When we ignore the Zeeman field, much of the vortex lattice energetics
of the lowest Landau level case remains valid with the Landau level mixing.
For instance, the two main formulas of Section V. A, Eqs.\eqref{EQ:freeScreen},
\begin{equation}
\langle \tilde{f} \rangle = -\tilde{\alpha} (C^2_{\uparrow\!\uparrow} + C^2_{\downarrow\!\downarrow}) +
\tilde{\beta}(C^4_{\uparrow\!\uparrow} + C^4_{\downarrow\!\downarrow}) + \tilde{\beta_3} C^2_{\uparrow\!\uparrow} C^2_{\downarrow\!\downarrow},
\end{equation}
and \eqref{EQ:freeFormulaLLL}),
\begin{equation}
\langle f \rangle = -\frac{H^2}{8\pi} - \frac{\tilde{\alpha}^2}
{2\tilde{\beta}+\tilde{\beta_3}},
\end{equation}
remains valid, mainly due to $|\hat{\Delta}| \propto (H_{c2} - H)^{1/2}$.
This means we can still calculate $h_s$, solely from quadratic terms. For
quadratic terms, we simply have two copies (for $s = \uparrow\!\uparrow$
and $\downarrow\!\downarrow$) of what was obtained for the case of
${\bf d} = (k_x + ik_y){\bf \hat{z}}$ by one of us
\cite{PhysRevB.58.14484}, we can use the formula for $h_s$ for
that case:
\begin{eqnarray}
h_s &=& \frac{8\pi^2 K}{\Phi_0} [C^2_{\uparrow\uparrow} \{(1-3\delta/\sqrt{2} + 2\delta^2)|\phi_0|^2\nonumber\\
&+& (2\delta^2 - \delta/\sqrt{2})|\phi_1|^2 + \delta^2|\phi_2|^2\}\nonumber\\
&+& C^2_{\downarrow\downarrow}\{(1-3\delta/\sqrt{2} + 2\delta^2)|\tilde{\phi}_0|^2\nonumber\\
&+& (2\delta^2 - \delta/\sqrt{2})|\tilde{\phi}_1|^2
+ \delta^2|\tilde{\phi}_2|^2\}].
\label{EQ:screenField2}
\end{eqnarray}
The spatial average of this equation is still proportional to
$(C^2_{\uparrow\uparrow}+ C^2_{\downarrow\downarrow})$ just like
Eq.~(\ref{EQ:screenField}). Also, $\tilde{\alpha} \propto (H_{c2} - H)$
still stands:
\begin{eqnarray}
\tilde{\alpha} &=& \frac{2\pi K (H_{c2} - H)}{\Phi_0} [(1-3\delta/\sqrt{2} + 2\delta^2)\langle|\phi_0|^2\rangle\nonumber\\
&+& (2\delta^2 - \delta/\sqrt{2})\langle|\phi_1|^2\rangle +
\delta^2\langle|\phi_2|^2\rangle].
\end{eqnarray}

However, the formula for $2\tilde{\beta}+\tilde{\beta_3}$
are much more complicated here, especially when we include
all terms of Eqs.~\eqref{EQ:int1} and \eqref{EQ:int2} for
these coefficients. For sake of convenience, instead of directly
writing down $\tilde{\beta}$ and $\tilde{\beta_3}$, we will list $\overline{h_s^2}$
(terms that are proportional to $K^2$), $\overline{f^{(4)}_{hom}}$
(terms involving coefficients of Eq.\eqref{EQ:int1}), and
$\overline{f^{(4)}_{in}}$ (terms involving coefficients of Eq.~\eqref{EQ:int2});
to obtain $2\tilde{\beta}+\tilde{\beta_3}$, we can use the relation
\begin{equation}
2\tilde{\beta}+\tilde{\beta_3} = \overline{f^{(4)}_{hom}} + \overline{f^{(4)}_{in}} - \frac{\overline{h_s^2}}{8\pi}.
\end{equation}
The following is the full listing of $\overline{h_s^2}/8\pi$, $\overline{f^{(4)}_{hom}}$, and $\overline{f^{(4)}_{in}}$ 
(note that we have set $\beta_3$ of Eq.~\eqref{EQ:int1} to be zero):
\begin{eqnarray}
\frac{\overline{h_s^2}}{8\pi} &=& \frac{8\pi^3 K^2}{\Phi_0^2} [(1-3\delta/\sqrt{2} + 2\delta^2)^2 \langle |\phi_0|^4 \rangle\nonumber\\
&+& 2 (2\delta^2 - \delta/\sqrt{2})(1-3\delta/\sqrt{2} + 2\delta^2)\langle |\phi_0|^2|\phi_1|^2 \rangle\nonumber\\
&+& 2\delta^2 (1-3\delta/\sqrt{2} + 2\delta^2)\langle |\phi_0|^2|\phi_2|^2 \rangle\nonumber\\
&+& (2\delta^2 - \delta/\sqrt{2})^2 \langle |\phi_1|^4 \rangle\nonumber\\
&+& 2\delta^2 (2\delta^2 - \delta/\sqrt{2}) \langle |\phi_1|^2|\phi_2|^2 \rangle + \delta^4 \langle |\phi_2|^4 \rangle\nonumber\\
&+& (1-3\delta/\sqrt{2} + 2\delta^2)^2 \langle |\phi_0|^2|\tilde{\phi}_0|^2 \rangle\nonumber\\
&+& 2 (2\delta^2 - \delta/\sqrt{2})(1-3\delta/\sqrt{2} + 2\delta^2)\langle |\phi_0|^2|\tilde{\phi}_1|^2 \rangle\nonumber\\
&+& 2\delta^2 (1-3\delta/\sqrt{2} + 2\delta^2)\langle |\phi_0|^2|\tilde{\phi}_2|^2 \rangle\nonumber\\
&+& (2\delta^2 - \delta/\sqrt{2})^2 \langle |\phi_1|^2|\tilde{\phi}_1|^2 \rangle\nonumber\\
&+& 2\delta^2 (2\delta^2 - \delta/\sqrt{2}) \langle |\phi_1|^2|\tilde{\phi}_2|^2 \rangle\nonumber\\
&+& \delta^4 \langle \langle |\phi_2|^2|\tilde{\phi}_2|^2 \rangle],\\
\overline{f^{(4)}_{hom}} &=& \beta_1 (\langle|\phi_0|^4\rangle + \delta^4 \langle |\phi_2|^4 \rangle) + 2\beta'_1 \delta^2 \langle|\phi_0|^2\rangle \langle |\phi_2|^2 \rangle\nonumber\\
&-& \beta_2 (\langle |\phi_0|^2|\tilde{\phi}_0|^2 + \delta^4
\langle |\phi_2|^2|\tilde{\phi}_2|^2 \rangle)\nonumber\\ &-&
2\delta^2 \beta'_2 \langle |\phi_0|^2|\tilde{\phi}_2|^2 \rangle,
\end{eqnarray}
and
\begin{eqnarray}
\overline{f^{(4)}_{in}} &=& \frac{2\gamma}{l^2} (\langle |\phi_0|^2|\tilde{\phi}_0|^2 \rangle - \langle |\phi_0|^2|\tilde{\phi}_1|^2 \rangle\nonumber\\
&+& 3\delta^2 \langle |\phi_0|^2|\tilde{\phi}_2|^2 \rangle - 3\delta^2 \langle |\phi_0|^2|\tilde{\phi}_3|^2 \rangle)\nonumber\\
&+& \frac{2\gamma'}{l^2}(3\delta^2 \langle |\phi_0|^2|\tilde{\phi}_2|^2 \rangle - 3\delta^2 \langle |\phi_0|^2|\tilde{\phi}_3|^2 \rangle\nonumber\\
&+& 2\delta^4 \langle |\phi_0|^2|\tilde{\phi}_1|^2 \rangle + \delta^4 \langle |\phi_0|^2|\tilde{\phi}_2|^2 \rangle - 3\delta^4 \langle |\phi_0|^2|\tilde{\phi}_3|^2 \rangle\nonumber\\
&+& 2\delta^4 \langle |\phi_1|^2|\tilde{\phi}_1|^2 \rangle + \delta^4 \langle |\phi_1|^2|\tilde{\phi}_2|^2 \rangle - 3\delta^4 \langle |\phi_1|^2|\tilde{\phi}_3|^2 \rangle\nonumber\\
&+& 3\delta^4 \langle |\phi_2|^2|\tilde{\phi}_2|^2 \rangle -
3\delta^4 \langle |\phi_2|^2|\tilde{\phi}_3|^2 \rangle).
\end{eqnarray}

\section{Correlation functions}
\label{app:corr} In calculating $\langle f^{(4)}_{in} \rangle$,
note
\begin{equation}
({\bm D} f)\cdot ({\bm D} g)^* = \frac{1}{l^2}[(\Pi_+ f)(\Pi_+
g)^* + (\Pi_- f)(\Pi_- g)^*]
\end{equation}
and
\begin{eqnarray}
\Pi_+ \phi_n &=& \sqrt{n+1}\phi_{n+1}\nonumber\\
\Pi_- \phi_n &=& \sqrt{n}\phi_{n-1}.
\end{eqnarray}
Together with partial integration
\begin{eqnarray}
\langle (\Pi_+ \phi_n) \tilde{\phi}^*_m \tilde{\phi}_p \phi^*_q
\rangle &=& \langle \phi_n (\Pi_-\tilde{\phi}_m)^* \tilde{\phi}_p
\phi^*_q \rangle\nonumber\\ &-& \langle \phi_n \tilde{\phi}^*_m
(\Pi_+\tilde{\phi}_p) \phi^*_q \rangle\nonumber\\ &+& \langle
\phi_n \tilde{\phi}_m^* \tilde{\phi}_p (\Pi_-\phi_q)^* \rangle,
\label{EQ:partialInt}
\end{eqnarray}
these equation gives
\begin{eqnarray}
\frac{1}{l^2}\langle\phi_0^*\tilde{\phi}_0({\bm D}
\phi_0)\cdot({\bm D} \tilde{\phi}_0)^* \rangle &=&
\langle |\phi_0|^2|\tilde{\phi}_0|^2 \rangle - \langle |\phi_0|^2|\tilde{\phi}_1|^2 \rangle\nonumber\\
\frac{1}{l^2}\langle\phi_0^*\tilde{\phi}_2({\bm D}
\phi_0)\cdot({\bm D} \tilde{\phi}_2)^* \rangle &=&
\langle |\phi_0|^2|\tilde{\phi}_2|^2 \rangle - \langle |\phi_0|^2|\tilde{\phi}_3|^2 \rangle\nonumber\\
\frac{1}{l^2}\langle\phi_2^*\tilde{\phi}_2({\bm D}
\phi_2)\cdot({\bm D} \tilde{\phi}_2)^* \rangle &=&
2 \langle |\phi_0|^2|\tilde{\phi}_1|^2 \rangle +  \langle |\phi_0|^2|\tilde{\phi}_2|^2 \rangle \nonumber\\
&+& 2 \langle |\phi_1|^2|\tilde{\phi}_1|^2 \rangle +  \langle |\phi_1|^2|\tilde{\phi}_2|^2 \rangle\nonumber\\
&-& 3 \langle |\phi_0|^2|\tilde{\phi}_3|^2 \rangle - 3 \langle |\phi_1|^2|\tilde{\phi}_3|^2 \rangle\nonumber\\
&+& 3 \langle |\phi_2|^2|\tilde{\phi}_2|^2 \rangle - 3 \langle |\phi_2|^2|\tilde{\phi}_3|^2 \rangle.\nonumber\\
\end{eqnarray}

These can be evaluated using
\begin{eqnarray}
\frac{\langle|\phi_p|^2|\phi_q|^2\rangle}{\langle|\phi_0|^2\rangle^2}  &=& \sum_{r,s} L^0_p({\bm k}_{rs}^2/2) L^0_q({\bm k}_{rs}^2/2) e^{-{\bm k}_{rs}^2/2},\nonumber\\
\frac{\langle|\phi_p|^2|\tilde{\phi}_q|^2\rangle}{\langle|\phi_0|^2\rangle^2} &=& \sum_{r,s} L^0_p({\bm k}_{rs}^2/2) L^0_q({\bm k}_{rs}^2/2) e^{-{\bm k}_{rs}^2/2}\cos({\bm k}_{rs}\cdot {\bm \tau})\nonumber\\
\end{eqnarray}
where $L^0_n$ is a Laguerre polynomial of $n$th order and ${\bm
k}_{rs} = (\sqrt{2\pi\sigma}r,\sqrt{2\pi/\sigma}(s-\varsigma r))$

%\section{Magnetic field}

%\bibliography{HQVlattice4b}

\end{document}